\documentclass[preprint,aps,showpacs,amsmath,eqsecnum]{revtex4}
\draft
\newfont{\cmu}{cmu10 scaled\magstep1}

\newcommand\PRD{{Phys. Rev.} D}

\newfam\BMath
\font\BMathL=cmmib10 
\font\BMathl=cmmib7
\font\BMathm=cmmib5
\textfont\BMath=\BMathL \scriptfont\BMath=\BMathl
\scriptscriptfont\BMath=\BMathm

\newcommand\E{\epsct{ e}}

\renewcommand\a{\alpha}
\renewcommand\b{\beta}
\newcommand\g{\gamma}

\newcommand\z{\zeta}
\newcommand\h{\eta}
\newcommand\q{\theta}

\renewcommand\k{\kappa}
\renewcommand\l{\lambda}
\newcommand\m{\mu}
\newcommand\n{\nu}

\newcommand\p{\pi}
\newcommand\s{\sigma}
\renewcommand\t{\tau}

\newcommand\f{\phi}

\renewcommand\o{\omega}

\def\mPi{{\mit\Pi}}			
\def\mTheta{{\mit\Theta}}		

\newcommand\D{\Delta}       
\renewcommand\L{\Lambda}      

\newcommand\ca{{\cal A}}

\newcommand\cm{{\cal M}}   
\newcommand\cn{{\cal N}}
   
\newcommand\cp{{\cal P}}   
\newcommand\cq{{\cal Q}}

\newcommand\cw{{\cal W}}

\def\ten #1{\bf#1}	    

\def\to{{\ten 0}}
\def\tA{{\ten A}}
\def\ta{{\ten a}}		
\def\tb{{\ten b}}		
\def\tB{{\ten B}}		
		
\def\tD{{\ten D}}

\def\tI{{\ten I}}		
\def\tv{{\ten v}}		
\def\tgv{{\ten u}}		
\def\tN{{\ten N}}               

\def\tM{{\ten M}}               
\def\tq{{\ten q}}               
\def\tP{{\ten P}}               
\def\tg{{\ten g}}
\def\gv{{\g\tv}}

\def\btD{\ten \D}

  \DeclareSymbolFont{UPM}{U}{eur}{m}{n}
  \SetSymbolFont{UPM}{bold}{U}{eur}{b}{n}
  \DeclareMathSymbol{\ubeta}{0}{UPM}{"0C}
  \DeclareMathSymbol{\ueta}{0}{UPM}{"11}
  \DeclareMathSymbol{\ugamma}{0}{UPM}{"0D}
  \DeclareMathSymbol{\uomega}{0}{UPM}{"21}
  \DeclareMathSymbol{\upartial}{0}{UPM}{"40}
  \DeclareMathSymbol{\upi}{0}{UPM}{"19}
  \DeclareMathSymbol{\usigma}{0}{UPM}{"1B}
  \DeclareMathSymbol{\utheta}{0}{UPM}{"12}
  \DeclareMathSymbol{\uvartheta}{0}{UPM}{"23}
  \DeclareMathSymbol{\ubtd}{0}{UPM}{"0E}

\def\aten{{\rm a}}		
\def\bten{{\rm b}}              
\def\Dten{{\rm D}}		
\def\gten{{\rm g}}		
\def\qten{{\rm q}}		
\def\uten{{\rm u}}		
\def\vten{{\rm v}}              

\def\tenG #1{\mbox{\boldmath$#1$}}	
\def\ten #1{\bf#1}			

\def\tnabla{{\tenG\upartial}}			
\def\tomega{{\tenG\uomega}}			
\def\tsigma{{\tenG\usigma}}			

\def\tD{{\tenG\nabla}}

\def\tpi{{\tenG\upi}}

\def\fractau#1{{\frac{1}{\tau_{#1}}}}

\newcommand\dd{\mbox{d}}   

\newcommand\Lra{\Longrightarrow}

\newcommand{\lan}{\langle}     
\newcommand{\ran}{\rangle}     

\newcommand{\3}{\frac{1}{3}}

\def\sks{,\;\;\;\;}

\newcommand\be{\begin{equation}}
\newcommand\ee{\end{equation}}
\newcommand\bea{\begin{eqnarray}}
\newcommand\eea{\end{eqnarray}}
\newcommand\beal{\begin{align}}
\newcommand\eeal{\end{align}}

\newcommand\ba{\begin{array}}
\newcommand\ea{\end{array}}
\newcommand\bc{\begin{center}}
\newcommand\ec{\end{center}}

\newcommand\bpi[1]{\begin{picture}#1}
\newcommand\epi{\end{picture}}

\def\jou#1#2#3#4{{#1} {\bf #2}, #3 (#4)}

\renewcommand\tv{{\tenG \vten}}
\renewcommand\tq{{\tenG \qten}}
\renewcommand\tgv{{\tenG \uten}}
\renewcommand\ta{{\tenG \aten}}
\renewcommand\tb{{\tenG \bten}}
\renewcommand\tg{{\tenG \gten}}

\newcommand{\PD}{{\partial}}
\newcommand{\btu}{\bigtriangleup}
\newcommand{\btd}{\bigtriangledown}

\newcommand{\ppr}{{\partial \over \partial r}} 
\newcommand{\epsct}[1]{\mbox{\boldmath${#1}$}}

\newcommand{\pr}{\prime}
\newcommand{\eps}{\varepsilon}

\newcommand{\lla}{\langle}
\newcommand{\gra}{\rangle}

\def\part{{\partial_t}}

\def\eqq{{\rm eq}}
\include{scrload}
\newfam\scrfam                                          
\global\font\twelvescr=rsfs10 scaled\magstep1%
\global\font\eightscr=rsfs7 scaled\magstep1%
\global\font\sixscr=rsfs5 scaled\magstep1%
\skewchar\twelvescr='177\skewchar\eightscr='177\skewchar\sixscr='177%
\textfont\scrfam=\twelvescr\scriptfont\scrfam=\eightscr
\scriptscriptfont\scrfam=\sixscr

\begin{document}


\title{ Relativistic Dynamics of Non-ideal Fluids: Viscous and heat-conducting
fluids \\
        I. General Aspects and 3+1 Formulation for Nuclear Collisions}
\author{Azwinndini Muronga$^{1,2}$}

\affiliation{$^1$Centre for Theoretical Physics \& Astrophysics, 
Department of Physics, University of Cape Town, Rondebosch 7701, South Africa\\
$^2$UCT--CERN Research Centre, 
Department of Physics, University of Cape Town, Rondebosch 7701, South Africa
}

\date{\today}

\begin{abstract}
Relativistic non-ideal fluid dynamics is formulated in 3+1 space--time
dimensions. The equations governing dissipative relativistic hydrodynamics are
given in terms of the time and the  3-space quantities which correspond to
those familiar from non-relativistic physics. Dissipation is accounted for by
applying the causal theory of relativistic dissipative fluid dynamics. As a
special case we consider a fluid without viscous/heat couplings in the causal
system of transport/relaxation equations. For the study of physical systems we
consider pure  (1+1)-dimensional expansion in planar geometry,
(1+1)-dimensional spherically symmetric ({\em fireball}) expansion,
(1+1)-dimensional cylindrically symmetric expansion and a  (2+1)-dimensional
expansion with cylindrical symmetry in the transverse  plane ({\em firebarrel}
expansion).  The  transport/relaxation equations are given in terms of the
spatial components of the dissipative fluxes, since these are not independent.
The choice for the independent components is analogous to the non-relativistic
equations.  

\end{abstract}

\maketitle
%
%
%
%
\section{Introduction}
%
%
The study of nonequilibrium properties of matter produced at relativistic high
energy nuclear collisons is important to unravel the underlying interactions
between the constituents of the system. Nonequilibrium effects are of central
importance to the space-time description of these collisions. 

In real fluids there are internal processes that result in transport of
momentum and energy from one fluid element to another on a microscopic level.
The momentum transport mechanisms give rise to internal frictional forces 
(viscous forces) that enter directly into the equations of motion, and that
also produce  frictional energy dissipation in the flow. The energy transport
mechanisms lead to energy conduction from one point in the flow to another.

In this work we use equations of fluid flow that explicitly account for the
processes described above. We adopt a continuum view here and leave the
microscopic kinetic-theory view to paper II of this work \cite{AMII}. In paper
II we recover essentially the same set of equations, but now with a much
clearer understanding of the underlying physics. The alternative approach also
allows us to evaluate explicitly, for a given molecular model, the transport
coefficients that are introduced on empirical grounds in the macroscopic
equations. For a review on
nonequilibrium models for relativistic heavy ion collisions see \cite{HS}.

To investigate the space-time evolution of high energy relativistic nuclear
collisions in a more complete  3+1 space-time formalism by means of dissipative
fluid dynamics requires numerical schemes to solve the fluid dynamical
equations of motion and the transport (evolution or relaxation) equations. The
existing numerical schemes for solving relativistic ideal fluid dynamics can be
used provided the relativistic dissipative fluid dynamics equations, in their
covariant structure, are cast in a more transparent form. 

The dynamics of relativistic non-ideal fluids is of considerable interest in
high energy heavy ion collisions \cite{AM2,MR,DT,AG,KP,PD,HiGy,HSC,TK,BRW,BR}.
In particular the recent application of causal relativistic fluid dynamics to 
heavy ion collisions \cite{MR} have proven to be excellent in explaining the
spectra of particles produced at the Relativistic Heavy Ion Collider (RHIC)
\cite{BRW,BR}.

The modeling of relativistic matter produced in these collisions is most
conveniently performed within the 3+1 formalism, where the 4-vectors, tensors,
and the equations of motion are decomposed  with respect to a general
space-time coordinate, allowing to express space-time derivatives in a more
transparent way as in non-relativistic mechanics.  The 3+1 representation of
the equations for relativistic ideal fluids been discussed by many authors (see
e.g. \cite{DHR}), in the context of applications to nuclear collisions.

Modeling dissipative processes requires non-equilibrium fluid dynamics and
irreversible thermodynamics. Standard dissipative fluid dynamics derived from
standard irreversible thermodynamics was first extended from non-relativistic
to relativistic fluids by Eckart \cite{CE}.  However, like in its
non-relativistic counterpart, in the Eckart theory and a variation thereof by
Landau and Lifshitz  \cite{LL},  dissipative fluctuations may propagate at an
infinite speed (causality problem). In addition, small perturbations of
equilibria driven by dissipative processes are unstable \cite{HL} and finally,
no well-posed initial value problem exists \cite{AM2}.  These problems in
standard dissipative fluid dynamics originate from the description of
non-equilibrium states via the local equilibrium states alone, i.e., it is
assumed that local thermodynamic equilibrium is established on an infinitely
short time-scale. This is a consequence of the assumption that the entropy
4--current includes only terms linear in dissipative quantities.  

Causal theories of dissipative fluids based on Grad's 14-moment method
\cite{HG} were formulated by M\"uller \cite{IM} and generalized
relativistically by Israel and Stewart \cite{IS}. They were formulated to
remedy some of the problems encountered in standard theories,  in particular
the causality problem. They are based upon extended irreversible
thermodynamics, where the entropy 4-current vector of standard thermodynamics
is extended by including terms that are quadratic in the dissipative
quantities. Hence the causal theory is also referred to as the second-order
theory of dissipative fluid dynamics. In causal theories of dissipative fluids
the set of thermodynamic variables is extended to include the dissipative
variables. The resulting transport equations and hence the equations of motion
are causal and have well--posed initial value problem for realistic equations
of state, transport coefficients, relaxation and coupling coefficients. Besides
the M\"uller-Israel-Stewart causal theory of relativistic non-ideal fluids
another causal relativistic theory for dissipative fluids exists, namely, a
theory by Liu {et al} \cite{LMR} which is of  divergence type, where the
dissipative fluxes are subject to a conservation equation. In this work we will
follow the former.
 
Causal theories of relativistic dissipative fluids provide the easiest
extension of standard theory towards finite signal speeds, and are
therefore appealing for nuclear collisions modeling. Extended fluid theories
are still rarely applied in the simulations of nuclear collisions,  which is
also due to the lack of an appropriate formulation. However there are
developments in this direction \cite{MR, HSC}. The purpose of this paper is
to provide such a formulation for dissipative relativistic fluid
dynamics.

This paper provides a complete set of equations for dissipative fluid dynamics
in their 3+1 dimensional representation, using a causal description of
thermodynamics.  Having at hand an appropriate 3+1 space-time formulation for
dissipative relativistic fluid theories, we can then employ them in modeling
relativistic nuclear collisions.  In section \ref{sec:basics} the basic
elements of causal dissipative hydrodynamics are presented in their covariant
compact form. In section \ref{sec:transp} we give the causal system of
transport equations in their covariant compact form. In section \ref{sec:wcl}
we give the weak coupling limit of the transport equations and the linearized
form of the complete system of the 14 equations. In section \ref{sec:3+1} we
give the  3+1 representation of relativistic dissipative fluid dynamics.   As
special cases for applications to the study of nuclear collision dynamics  we
specify the system of equations given in Section \ref{sec:3+1} to the case of
(i) (1+1)--dimensional  expansion in planar geometry, (ii)
(1+1)-dimensional cylindrically symmetric expansion and (iii)
(2+1)-dimensional expansion with cylindrical symmetry in the transverse plane
(firebarrel expansion) in Section \ref{sec:physprobs}. The (1+1)-dimensional
spherically symmetric expansion is left for  Appendix \ref{append:sphsym}.

Throughout this article we adopt the units $\hbar = c = k_{\rm B} = 1$. The
signature of the metric tensor is always taken to be $g^{\mu \nu} = \mbox{diag}
(+1,-1,-1,-1)$.   Upper greek indices are contravariant, lower greek indices
covariant. The greek indices used in 4--vectors go from 0 to 3 ($t,x,y,z$) and
the roman indices  used in 3--vectors go from 1 to 3 ($x,y,z$).  The scalar
product of two 4-vectors $a^\mu\, , \,\, b^\mu$ is denoted by  $a^\mu\, g_{\mu
\nu}\, b^\nu \equiv a^\mu \, b_\mu $. The scalar product of two 3-vectors is
denoted by bold type, namely,  $\ta,\, \tb$,  $\ta \cdot \tb$. The notations
$A^{(\a\b)} \equiv  (A^{\a\b}+A^{\b\a})/2$ and $A^{[\a\b]} \equiv 
(A^{\a\b}-A^{\b\a})/2$ denote symmetrization and antisymmetrization
respectively. The 4--derivative is denoted by $
\PD_\a \equiv \PD/\PD x^\a$. Dot
product between two tensors will be denoted $A^{\m\n} B_{\m\n}$ while for
spatial components of the tensor $\tA\cdot \tB \equiv \sum_k A^{ik} B^{kj}$ and
for the double scalar product between tensors $\tA:\tB \equiv \sum_{i,k} A^{ik}
B^{ki}$. The notation $\tb\otimes \tb = b^i b^j$ and similarly  $\tb\otimes
(\tB\cdot\ta) = b^i \sum_k B^{jk}a^k$ denotes the multiplication of two
vectors.

%
%

%
\section{Non-ideal relativistic fluid dynamics} 
\label{sec:basics}

Non-ideal fluids, like ideal fluids, are described by conservation laws for
the particle current vector and the energy-momentum stress tensor. A
thermodynamic equilibrium state of a non-ideal fluid dynamics theory can be
characterized by the five independent dynamical fields which describe a
reversible thermodynamic process in an ideal fluid dynamics, and one often
chooses two scalar equilibrium state variables and the three independent
components of the preferred 4-velocity. In high energy heavy-ion collisions we
generally take the net baryon number density and energy density as state
variables. A general non-equilibrium state or irreversible thermodynamic
process is to be characterized by 14 independent dynamical fields.
These are, in addition to the ideal fields, the projections of the
stress-energy tensor parallel and orthogonal to the preferred velocity
(dissipative fluxes), e.g. the trace and trace-free part of the spatial stress
tensor, and the spatial heat flux vector.

We choose the average 4-velocity $u^\m$\/ such that the particle flux in the
associated rest frame vanishes. This is the Eckart frame \cite{CE} and is the 
natural frame in systems where there exists  some conserved net charge (see
\cite{IS} for the alternative energy frame description). The state of the fluid
is assumed close to a thermodynamic equilibrium state, characterized by the
local thermodynamic equilibrium scalars such as the equilibrium net  charge
density $n_{\eqq}$, equilibrium energy density $\eps_{\eqq}$ the local
equilibrium velocity $u^\m_{\eqq}$ which in the Eckart frame can be chosen such
that only the pressure $p$\/ deviates from the local equilibrium pressure
$p_{\eqq}$\/ by the bulk viscous pressure $\mPi=p-p_{\eqq}$\/, whereas
$n=n_{\eqq}$\/ and  $\eps=\eps_{\eqq}$\/. Other thermodynamic quantities such
as the temperature, chemical potential and entropy are obtained from the
equation of state and thermodynamic relations.

We consider a fluid that consists of a single component. The variables of
concern are the net charge 4-current $N^{\mu}$, the energy-momentum stress
tensor $T^{\mu\nu}$ and the entropy flux $S^{\mu}$. The divergence of
energy--momentum tensor and of the net charge 4--current vanishes locally. That
is, the energy--momentum and net charge are conserved locally. However in
general the  divergence of entropy 4--current does not vanish. The second law
of thermodynamics requires that it be a positive and nondecreasing function. In
equilibrium the entropy is maximum and the divergence of the entropy 4--current
vanishes. Thus 

\bea
\PD_{\mu}N^{\mu} & = & 0 \label{eq:Ncons}~, \\
\PD_{\nu}T^{\mu\nu} &= &0 \label{eq:EMcons}~,\\
\PD_{\mu}S^{\mu}  &\geq&  0 \label{eq:2ndLaw}~,
\eea
where 
\bea
N^{\mu} &= & n u^{\mu} ~,\label{eq:4-number}\\
T^{\mu\nu} &=& (\eps +p+\mPi)u^{\mu}u^{\nu} - (p+\mPi) g^{\mu\nu} 
               + 2 q^{(\m} u^{\n)}
               + \pi^{\mu\nu}  ~, \label{eq:tmunu}\\
S^\mu &=& s u^\mu + \b q^\mu 
          -{1\over2}\b u^\mu \biggl(\beta_0 \mPi^2
- \beta_1 q^\n q_\n +\beta_2 \pi^{\lambda\nu}\pi_{\lambda\nu}\biggr) \nonumber
\\
&&~~~~~~~~~~~~~~~~~~~~~~~~~
-\b \biggl(\alpha_0 q^\mu \mPi 
- \alpha_1 q_\nu \pi^{\mu\nu}\biggr)~.
 \label{eq:4-entro}
\eea

In the local rest frame defined by $u^\mu=(1,\bf{0})$ the quantities appearing
in the decomposed tensors take their  actual meanings:
$n  \equiv u_\mu N^\mu$ is the net charge density,
$\eps \equiv u_\mu T^{\mu\nu} u_\nu$ is the energy density,
$p + \mPi \equiv -\frac{1}{3} \btu_{\mu\nu} T^{\mu\nu}$ is the local isotropic 
pressure plus bulk pressure,
$q^\mu \equiv u_\nu T^{\nu\lambda} \btu^\mu_\lambda $
is the heat flow, 
$\pi^{\mu\nu} \equiv T^{\langle\mu\nu\rangle}$ is the shear stress
tensor, and 
$s \equiv u_\mu\, S^\mu $ is the entropy density.  
The angular bracket notation, representing the symmetrized spatial and
traceless part of the tensor, is defined by 
$A^{\lla\mu\nu\gra} \equiv \left[\frac{1}{2}\left(\btu^\mu_\sigma
\btu^\nu_\tau
+\btu^\mu_\tau \btu^\nu_\sigma\right)
-\frac{1}{3}\btu^{\mu\nu}\btu_{\sigma\tau}\right]
A^{\sigma\tau} $.
The space--time derivative decomposes into 
$ \PD^\mu = u^\mu D + \btd^\mu$ with $u^\mu \btd_\mu=0 $. 
In this space--time derivative decomposition $D \equiv
u^\mu\PD_\mu $ is the convective time derivative and  
$\btd^\mu \equiv \btu^{\mu\nu}\PD_\nu $ is the gradient operator.
The projection onto the 3-space  
$\btu^{\mu\nu} \equiv g^{\mu\nu}-u^\mu u^\nu \equiv \btu^{\nu \mu}$ is orthogonal to 
$u^\mu$, that is, $\btu^{\mu \nu} u_\nu = 0$ 
and $u^{\mu}$ is the hydrodynamical 4-velocity of the net charge and is to be normalized such
that \begin{math} u^\mu u_\mu =1 \end{math} and therefore \begin{math} u^\mu \PD_\nu
u_\mu =0 \end{math}. 
Here $\b\equiv
1/T$ is the inverse temperature. 
The $\a_i(\eps,n)$ and $\b_i(\eps,n)$ in (\ref{eq:4-entro}) are the second
order coefficients which are expressed in terms of  thermodynamic integrals and
therefore are given by the equation of state. These second order coefficients
are presented in more details in paper II of this work \cite{AMII}.

From the entropy 4-current (\ref{eq:4-entro}) one sees that the entropy density
and flux are respectively given by
\bea
s&=&u_\m S^\m = s(\eps,n) - {1\over 2}\b \biggl(\b_0 \mPi^2 - \b_1 q_\n q^\n 
+ \b_2 \pi_{\l\n}\pi^{\l\n}\biggr) ~, \label{eq:sdens}\\
\Phi^\m &=& \btu^{\m\n} S_\n =  \b q^\mu  - 
             \b (\a_0 \mPi q^\m - \a_1 \pi^{\m\n}q_\n) \label{eq:sflux}
~.
\eea 
Note that the entropy density is independent of the $\a_i$ while the entropy
flux is independent of the $\b_i$. 
The negative sign of the non-equilibrium contributions reflects the fact that
the entropy density is maximum in equilibrium. 
The thermodynamic coefficients $\beta_i(\eps,n)\ge0$\/ in (\ref{eq:sdens}) model 
deviations of the physical entropy density from $s$\/ due to scalar/vector/tensor dissipative 
contributions to $S^\m$\/. 
The $\alpha_i(\eps,n)$\/ in (\ref{eq:sflux}) model contributions due to viscous/heat coupling,
which do not influence the physical entropy density.

\section{Relaxation transport equations}
\label{sec:transp}

As we mentioned before we shall use the M\"{u}ller-Israel-Stewart second order
phenomenological theory for dissipative fluids \cite{IM,IS}. Essentially the
extended irreversible thermodynamics theory rests on two hypothesis: (1) The
dissipative flows, heat flow and viscous pressures, are considered as
independent variables. Hence the entropy function depends not only on the
classical variables, net baryon density and energy density, but on these
dissipative flows as well. (2) At equilibrium, the entropy function is maximum.
Moreover, its flow depends on all dissipative flows and its production rate is
semi-positive definite. As a consequence the bulk pressure $\mPi$, the heat
flow $q^\m$, and the traceless shear viscous tensor $\pi^{\m\n}$ obey the
evolution equations  \cite{IS}
\bea
\tau _\mPi \dot{\mPi }+\mPi &=& -\zeta \mTheta 
-\frac 12\zeta T \mPi \PD_\m\left( \frac{\tau_\mPi u^\m }{\zeta T}\right)  
+l_{\mPi q}\btd_\m q^\m ~,\label{eq:bulk} \\
\tau _q \btu_\nu ^\mu \dot{q}^\nu + q^\mu &=& \kappa T \biggl({\btd^\m
T\over T}-a^\m\biggr)
+ \frac 12\kappa T^2 q^\mu\PD_\n\left( \frac{\tau _q u^\n }{\kappa T^2}\right)   \nonumber\\
&&-l_{q\pi}\btd_\nu\pi^{\mu\nu} -l_{q\mPi}\btd^\mu\mPi 
+\tau _q \omega ^{\mu\nu } q_\nu ~,\label{eq:heat} \\
\tau _\pi \btu^{\m \a} \btu^{\n\b} \dot{\pi }_{\alpha \beta }+\pi ^{\mu\nu} 
&=& 2\eta \s^{\m\n}-\frac 12\eta T \pi ^{\mu \nu }\PD_\l\left(
\frac{\tau _\pi
u^\l }{\eta T}\right) \nonumber \\
&&+l_{\pi q}\btd^{\lan\m}q^{\n\ran}
+2\tau _\pi \pi ^{\a(\mu} \omega ^{\nu )}_\alpha~,\label{eq:shear}
\eea
where $\omega ^{\mu \nu }=\btu^{\mu \a} \btu^{\nu \b} \PD_{[\b} u_{\alpha]}$
is the vorticity, $\mTheta = \btd_\m u^\m $ is the expansion scalar,
$\sigma^{\m\n} = \btd^{\lan\m}u^{\n\ran}=\btd^{(\m} u^{\n)}
 -\frac 13\btu^{\m \n }\btd_\l u^\l $ is the shear tensor and $a^\m \equiv
 \dot{u}^\m$ is the acceleration 4-vector. The local enthalpy 
density is $w = \eps+p$. 
Overdot denotes $\dot{A}_{\alpha
\beta }=u^\l \PD_\l A_{\alpha \beta}$. The transport coefficients   
$\kappa $, $\zeta $, and $\eta $ denote the thermal conductivity, and the
bulk and shear viscous coefficients respectively. The quantities 
\bea
\t_\mPi &=& \zeta \b_0 \sks \t_q =\kappa T \b_1 \sks \t_\pi = 2\eta \b_2 
~, \label{eq:relt}\\
l_{\mPi q}&=& \zeta \a_0 \sks l_{q\mPi} =\kappa T\a_0 \sks
l_{q\pi} = \kappa T \a_1 \sks l_{\pi q} = 2\eta \a_1 ~, \label{eq:rell}
\eea
 are the relaxation times for the bulk pressure  $\tau _\mPi $,  the heat flux
$\tau _q $ and the shear tensor $\tau _\pi $ ; and the relaxation lengths for
coupling between heat flux and bulk pressure ($l_{\mPi q}$, $l_{q\mPi}$) and
between heat flux and shear tensor ($l_{q\pi}$, $l_{\pi q}$).  The $\a_i$ and
$\b_i$  are presented in Ref.~\cite{AMII}. 

%
\section{The weak coupling limit of causal transport system}
\label{sec:wcl}

The complexity of the full evolution equations (\ref{eq:bulk})-(\ref{eq:shear})
makes their applications tractable only if certain simplifications are made.  A
particularly simple set of evolution equations results from the    assumption
that there is no viscous/heat coupling (specifically $\a_0 =\a_1 =0$ ). We will
also drop the term with the one--half factor by assuming that the thermodynamic
gradients are small.  Finally we make a further assumption that there are no
coupling  between acceleration and dissipative fluxes. Such is the case in
systems where the gradients in thermodynamic quantities are small and in
systems where there is no coupling of velocity components. In relativistic
heavy ion collisions such cases are, for example, the scaling solution
\cite{Bjorken}, 
effective (1+1) dimensional expansions (expansions in planar geometry, in
cylindrical/spherical geometry with symmetry -- pure radial expansions). The
evolution equations resulting from (\ref{eq:bulk})-(\ref{eq:shear}) under  the
above assumptions are 
\bea
\dot{\mPi} &=& \fractau{\mPi}\Big(\mPi_E-\mPi\Big) \label{eq:Pi} ~,\\
\dot{ q}^\m &=&	\fractau{q}\Big(q^\m_E-q^\m\Big)   \label{eq:q} ~,\\
\dot{\pi}_{\m\n} &=& \fractau{\pi}\Big(\pi^{\m\n}_E-\pi^{\m\n}\Big)
\label{eq:pi} ~,
\eea
where  
\bea
\mPi_E &=& -\zeta \mTheta  \label{eq:PiE}~,\\
q^\mu_E &=& \k T 
\left(\frac{\btd^\mu T}{T} - \dot{u}^\mu \right) \label{eq:qE}~,\\
\pi^{\mu\nu}_E &=& 2\eta \s^{\mu\nu} \label{eq:piE}\enspace ,
\eea
are the standard Eckart thermodynamic fluxes.  Equations
(\ref{eq:Pi})-(\ref{eq:pi}) are of covariant relativistic 
Maxwell-Cattaneo form. 
In contrast to the algebraic constraint equations
(\ref{eq:PiE})-(\ref{eq:piE}), the evolution equations
(\ref{eq:Pi})-(\ref{eq:pi}) are first order partial  differential equations,
which assure that in the local rest frame the viscous bulk/shear stresses and
the heat flux relax towards their standard limits
$\{\mPi_E\,,q^\m_E\,,\pi^{\m\n}_E\}$\/  on time-scales $\tau_A$\/. The
relaxation times $\tau_A$\/ follow in principle from kinetic theory \cite
{AMII}.

The conservation laws, that is, the net charge conservation equation, 
the equations of motion, and the energy
equation together with the
evolution equations (\ref{eq:Pi})--(\ref{eq:pi}) constitute a complete
system  of hyperbolic first order partial differential equations for the
solution vector of 14 (=1+1+3+1+3+5) dynamical variables  $\{n$\/, $\eps$\/,
$u^\m$\/, $\mPi$\/, $q^\m$\/, $\pi^{\m\n}\}$\/. This system represents a
14-Fields Theory for relativistic dissipative fluids which is causal and stable
for appropriate equation of state, initial conditions, transport and relaxation
coefficients \cite{AM2}.

The equations of non-ideal fluid dynamics and in particular those of
relativistic dissipative fluid dynamics are in general rather complicated.
It is therefore useful to have a simple means for judging both the relative
importance of various phenomena that occur in a flow, and the flow's
qualitative nature. This is most easily done in terms of a set of dimensionless
numbers which provide a convenient characterization of the dominant physical
processes in the flow. Flows whose physical properties are such that they
produce the same values of these numbers can be expected to be qualitatively
similar even though the value of any one quantity --say velocity or
characteristic length/time --may be substantially different from one flow to
another. Thus the assumption made in this section (the weak coupling limit)
must be thoroughly checked on a case by case  for the problem under
consideration. This is the subject of current investigation and will be
presented somewhere.

The analysis of the instantaneous dynamics of a relativistic fluid can be
summarized by the identity  
\be
\PD^\n u^\m = a^\m u^\n + \s^{\m\n} + \o^{\n\n}+ {1\over 3} \theta \btu^{\m\n}
~, \label{eq:kinematics}
\ee
which is a generalization of the Cauchy-Stokes decomposition theorem.  It shows
that at each space-time point a fluid is accelerated along its proper time
axis, and experiences shear, rotation (vorticity), and expansion along its
local space axes. Explicit expressions for $\btu^{\m\n}$, $a^\m$, $\theta$,
$\o^{\m\n}$, and $\s^{\m\n}$ in terms of the ordinary velocity $\tv$ and
lab-frame space and time derivatives are given in section \ref{sec:3+1}.   Note
that from $u_\m \btu^{\m\n} =0$ we have  $u_\m\s^{\m\n}=u^\m \s_{\m\n}=0$  and
$u_\m\o^{\m\n}=u^\m \o_{\m\n}=0$  and from $\btu^{\m\n} \equiv
\btu^{\m\a}\btu^\n_\a$ and  $\btu^{\m\n}\btu_{\m\n} =3$ we have 
$\btu_{\m\n}\s^{\m\n}=\btu^{\m\n}\s_{\m\n}=0$.  From the shear tensor we define
a shear scalar  $\s$ from
\be
\s_{\m\n}\s^{\m\n}  =  {1\over 6}\s ~.
\ee
%

\section{Relativistic dissipative fluid dynamics in 3+1 formulation}   
\label{sec:3+1}             
\subsection{3+1 space--time formulation}

In order to find a numerical solution to the relativistic dissipative fluid
dynamics equations, we must first cast the equations into suitable form. The
equations are usually written in a more compact covariant form. The equations
in this form are not well suited for solving on a computer. First it is unclear
what information to specify on the boundary of the 4 dimensional space-time in
order to get a well posed problem. Second, this form of the equations does not
allow us to pose the type of physical questions we often wish to ask. Typically
we want to be able to specify a physical system at some initial time and find
the development of this system as time evolves. Because of these two issues, we
break the covariance of the equations. The covariance is broken in two steps:
(i) First we split the 4 dimensional  space-time coordinates into 3 dimensional
spatial coordinates plus 1 dimensional  time coordinate. (ii)
Secondly, we break a 2nd-rank tensor up into 1+3+6 $(00,0i,ij)$ independent
components. A 4-vector is broken into 1+3 $(0,i)$ components while a scalar
is unchanged. The numerical problem of constructing the 3+1 space-time is (i)
Determine a set of initial conditions that satisfy conservation equations and
transport equations. (ii) Evolve this set of initial conditions forward in
time.

In this section we will give explicit expressions for the net charge 4-current 
$N^\m$, energy-momentum stress tensor $T^{\m\n}$, net charge conservation
$\PD_\m N^\m=0$, energy-momentum conservation $\PD_\m T^{\m\n}$ in terms of the
ordinary velocity and lab frame space and time derivatives. In addition we
also give the explicit expressions for the transport/relaxation equations for
the dissipative quantities, namely the bulk pressure equation, the heat flux
equation and the shear stress tensor equation in terms of the ordinary
velocity and lab frame space and time derivatives.

The 4-velocity, metric tensor and projection tensor are decomposed as 
\bea
u^\m &=& (\g,\tgv) \sks \tgv =\gv \sks \g = (1-\tv\cdot\tv)^{1/2}\,\\
g^{\m\n} &=& (1,\to, \tg) \sks \to \equiv \{g^{0i}\} 
            \sks \tg \equiv \{g^{ij}\} =\begin{cases}
	                                     -1& \text{if $i=j$}~,\\
					      0& \text{if $i \neq j$}~,
					      \end{cases}\\
\D^{\m\n} &=& -(\g^2 v^2, \g\tgv, -\btD) 
            \sks \btD = \tg-\tgv\otimes\tgv
\eea
where $\gamma = (1-\tv^2)^{-1/2}$ and $\tv$ is the three velocity of the fluid. 
The space-time gradient is decomposed as  
\bea
\partial^\m &=& (\part,\tnabla) \equiv u^\m D +\btd^\m\,,\\
D &=& (\g\part,\Dten) \sks \Dten = \tgv\cdot\tnabla 
\sks \tnabla=\{\partial_i\} ~,\\
\btd^\m &=& -(\g^2 v^2\part+\g\tgv\cdot{\tnabla}, \g\tgv\part-\tD)
\sks \tD = \btD\cdot\tnabla\,,
\eea
In the irreducible decomposition
\bea
\tnabla \otimes \tgv   &=&     \tsigma
                        +\tomega
                        +\frac{1}{3}\vartheta\ten \btD
                        + \ta\otimes\tgv
\label{eqDecompAS}\;,
\eea
the kinematic properties of the fluid, i.e, expansion scalar, acceleration, shear and
vorticity tensors, are given by 
\bea
\mTheta &=& \part \g +\tnabla\cdot \tgv \,,\label{eq:volexp1}\\
\ta &=& \g \part \tgv + \Dten\tgv \,,\label{eq:acc1}\\
\tsigma &=& ^(\tnabla\otimes\tgv^) - ^(\ta\otimes\tgv ^)
-\frac{1}{3}\mTheta\btD \,,\label{eq:sigm1}\\
\tomega &=& ^[\tnabla \tgv^] + ^[\ta\otimes \tgv^] ~.
\label{eq:vort1}
\eea
These kinematic quantities can be written in terms of time derivatives and
spatial ones
\bea
\mTheta &=& \part \g + \vartheta \,,\label{eq:volexp2}\\
\ta &=& \g \part \tgv + \ta_s \,,\label{eq:acc2}\\
\tsigma &=& -\{\g ^(\tgv \otimes \part \tgv^) +\3\btD\part\g  -\tsigma_s\} \,,\label{eq:sigm}\\
\tomega &=& \g ^[\tgv\part\otimes \tgv^] + \tomega_s ~,
\label{eq:vort2}
\eea
where
\bea
\tsigma_s &\equiv&  ^(\tnabla\otimes \tgv^) -
^(\ta_s\otimes\tgv^)-\3\btD\vartheta ~,\\
\tomega_s &\equiv& ^[\tnabla \otimes\tgv^] + ^[\ta_s\otimes\tgv^] ~,\\
\ta_s &\equiv& \Dten\tgv ~,\\
\vartheta &\equiv& \tnabla\cdot\tgv ~.
\eea

A complete split involves 3+1 representations for the fields
$N^\m$, $T^{\m\n}$, and $S^\m$, together with the equations governing their
evolution. The representation of $T^{\m\n}$ and $S^\m$ involves the 3+1
formulation of dissipative quantities, i.e., $\mPi$, $q^0$ and $\tq =\{q^i\}$,
as well as $\pi^{00} \,,\pi^{0i}$ and $\tpi = \pi^{ij}$ of dissipative fluxes
$(\mPi\,,q^\m\,,\pi^{\m\n})$. Not all of them are independent. Orthogonality to
the fluid velocity field, $q^\m u_\m = \pi^{\m\n} u_\n =0$, together with
$\pi^\n_\n=0$, yields three and five independent components of heat flux and
shear tensor, respectively. Analogy with non-relativistic physics suggests to
generally eliminate $q^0$ in favor of $\tq$, as well as $\pi^{00}$ and
$\pi^{0i}$ in favor of $\tpi$, though different choices might be more
appropriate in particular applications. For discussion on different choices of
the dissipative fluxes see \cite{HSC}. Orthogonality also naturally holds for
dissipative forces $\btd^\m \ln T - a^\m$ and $\s^{\m\n}$. The 3+1
representation of the orthogonality relations for $q^\m$ and $\pi^{\m\n}$ is
\bea
0&=& \g (q^0  - \tq\cdot\tv)  \Longrightarrow  q^0=\tq\cdot\tv \,, \label{qu-eq}\\
0&=& \g(\pi^{00}  - \pi^{0i}\cdot\tv)   \Longrightarrow 
\pi^{00}=\pi^{0i}\cdot\tv=(\tpi\cdot\tv)\cdot\tv \,, \label{piu0-eq}\\
\ten{0}&=& \g(\pi^{0i} -\tpi\cdot\tv)   \Longrightarrow 
           \pi^{0i}=\tpi\cdot\tv \,, \label{piu1-eq}
\eea
where Eq.~(\ref{piu1-eq}) was used in the last equality of Eq.~(\ref{piu0-eq}). 
Finally, $\pi^\n_\n =0$ becomes
\be
\pi^{00} - \mbox{tr}(\tpi) =0 \Longrightarrow 
                   \mbox{tr}(\tpi) = (\tpi\cdot\tv)\cdot\tv \,, \label{pipi-eq}
\ee
and allows one to eliminate one component of $\tpi$. Substituting
$q^0\,,\pi^{00}$ and $\pi^{0i}$ in this way, it remains to provide expressions
for one scalar function, the bulk viscous pressure $\mPi$, three components of
the  heat flux (i.e, ${\tq}$), and five components of the viscous stress tensor
(i.e., ${\tpi}$), as in non-relativistic physics.

\subsection{Conservation Laws}
\label{sec:conserv}

The local net charge density $N^0$ is subject to the 3+1 representation of
Eq.~(\ref{eq:Ncons}), and the 3+1 representation of Eq.~(\ref{eq:EMcons}) yields
conservation laws both for the local total energy density $T^{00}$ and the 
local momentum density $T^{0i}$,
\bea
\PD_\m N^\m &\equiv& \part \cn + \tnabla\cdot \left\{\cn \tv\right\}
=0\,,\label{eq:Nmcons}\\
\PD_\m T^{\m 0} &\equiv& \part E +\tnabla\cdot \left\{(E+\cp)\tv + \g(\tq-(\tq\cdot\tv)\tv)   
            + \tpi\cdot\tv -((\tpi\cdot\tv)\cdot\tv)\tv\right\}=0\,,\label{eq: Econs}\\
\PD_\m T^{\m i}&\equiv& \part \tM +\tnabla\cdot \left\{\tM\otimes\tv-\cp\tg
         +\g(\tv\otimes\tq-(\tq\cdot\tv)(\tv\otimes\tv))
	+\tpi-(\tpi\cdot\tv)\otimes\tv\right\} = {\ten{0}}\,,\label{eq:Mcons}
\eea
Thus the charge 4-current vector $N^\m$ and the stress-energy tensor $T^{\m\n}$ are
represented by $N^\m \equiv (\cn,~\tN)$ and $T^{\m\n} \equiv (E,~\tM,~\tP)$ 
where 
\bea
\cn \equiv N^0 &=& n \g \,,\label{eq:cn}\\
\tN \equiv n \g\tv &=& \cn\tv \,,\label{eq:tn}\\
E\equiv T^{00} &=& (\eps+\cp)\g^2 - \cp + 2\g\tq\cdot\tv
+(\tpi\cdot\tv)\cdot\tv \,,\label{eq:E}\\
\tM\equiv \{T^{0i}\}_{i=x,y,z} &=& (\eps+\cp)\g^2\tv +\g(\tq +(\tq\cdot\tv)\tv) 
+\tpi\cdot\tv \,,\label{eq:M}\\
\tP\equiv \{T^{ij}\}_{i,j=x,y,z} &=& (\eps+\cp)\g^2\tv\otimes\tv - \cp\tg 
+2\g ^{(} \tv\otimes\tq ^{)} +\tpi \,,\label{eq:P}
\eea
with $\cp=(p+\mPi)$ the effective pressure.  The physical meaning of the
components of the energy-momentum stress tensor is as follows:  $T^{00}$ is the
total energy density of the fluid,  $T^{0i}$ is the energy flux density in the
$i$th direction of the flow,  $T^{i0}$ is the momentum density in the $i$th
direction,  $T^{ij}$ is the rate of transport of the $i$th component of the
momentum per unit volume through a unit area oriented perpendicular to the
$j$th coordinate axis.   For a fluid at rest every element of area experiences
only a force normal to the surface of the element and this force is independent
of the orientation of the element. If the fluid is ideal, then it is
non-viscous and will not support tangential stress even when the fluid is in
motion. The stress acting across a surface in an ideal fluid is thus always
normal to the surface. A non-ideal fluid will not support tangential stress
when it is at rest, but can do so when it is in motion. The energy-momentum
stress tensor has the {\em viscous stress tensor}. The pressure tensor $T^{ij}$
gives the normal component of the surface force (in the $j$ direction) acting
on a surface element that is oriented perpendicular to the $i$th coordinate
axis. The spatial diagonal components of $T^{\m\n}$ are  the normal
stresses while the off-diagonal spatial components are the tangential stresses
(or shearing stresses).

From Eqs.~(\ref{eq:E})-(\ref{eq:P}) we can write $\tM$ and $\tP$ as
\bea
\tM &=& (E+\cp)\tv + \g(\tq-(\tq\cdot\tv)\tv) +
\tpi\cdot\tv-((\tpi\cdot\tv)\cdot\tv)\tv \,,\label{eq:tM}\\
\tP &=& \tM\otimes\tv - \cp\tg +\g(\tv\otimes\tq - (\tq\cdot\tv)(\tv\otimes\tv))
+\tpi - (\tpi\cdot\tv)\otimes\tv \,, \label{eq:tP}
\eea
and the local net charge density and energy density are found from
\bea
\eps &=& E - (\tM+\g^{-1}\tq)\cdot\tv \,, \label{eq:eps}\\
n &=& (1-\tv\cdot\tv)^{1/2}\cn \,\label{eq:n}.
\eea
In ideal fluids one notices that $\tM$ and $\tv$ are parallel and one can find
$\tv$ from the simple expressions \cite{DHR}. In non-ideal fluids $\tM$ and
$\tv$ are in general not parallel. Only in special cases where the fluid
velocity components decouple can one still finds $\tM$ and $\tv$ parallel and
one can still solve for velocity \cite{MR}. In the case where there is strong
coupling between the velocity components one has to solve for $\tv$ iteratively
from Eqs.~(\ref{eq:tM}) or from from Eqs.~(\ref{eq:eps}) and (\ref{eq:n}).

\subsection{Maxwell-Cateneo limit of the  transport equations in the 3+1 
formalism}
\label{sec:weak}

Due to orthogonality it suffices to provide transport equations only for
spatial components of the dissipative fluxes, i.e.  $\{\mPi\,,\tq\,,\tpi\}$.
For particular set, Eqs.~(\ref{eq:Pi})-(\ref{eq:pi}), this yields
\bea
\g \part \mPi +\gv\cdot\tnabla\mPi &=&\frac{1}{\tau_\mPi} (\mPi_E -\mPi)
\,,\label{eq:mcbulk} \\
\g \part \tq +\gv\cdot\tnabla \tq &=& \frac{1}{\tau_q} ({\tq}_E-\tq) ~,
\label{eq:mcheat}\\
\g \part \tpi +\gv\cdot\tnabla \tpi &=&\frac{1}{\tau_\pi} (\tpi_E -\tpi)
~.\label{eq:mcshear}
\eea
The complete 3+1 representation of the constraints, 
Eqs.~(\ref{eq:PiE})-(\ref{eq:piE}) can be written down with the help of the
orthogonality relations Eqs.~(\ref{qu-eq})-(\ref{pipi-eq})
\bea
\mPi_E &=& -\zeta\mTheta \,, \\
{\tq}_E &=& -\lambda T(\g^2\tv\part \ln T -\tD \ln T + \ta) \,,\\
\tpi_E &=& 2\eta \tsigma ~. \
\eea
Note that the weak coupling limit of the causal transport equations can be further simplified for
special cases such as the scaling solution case. In that case the last term
(acceleration coupled to heat or shear flux) on the right hand side of
Eqs.~(\ref{eq:mcheat})and (\ref{eq:mcshear}) vanishes. This is also the case in
simple (1+1) dimensional problems as we will see in the following section(s).

\section{Physical Problems}
\label{sec:physprobs}
In this section we consider special cases with emphasis on different
geometries (coordinate systems). These cases represent some of the various scenarios one can find in
high energy nuclear collisions. In this regard the longitudinal direction is
taken to be the $z$ axis in all scenarios. We will consider: a pure
(1+1)-dimensional expansion along the longitudinal $z$ axis in planar geometry,
a (1+1)-dimensional cylindrically symmetric expansion along the radial direction
($r$ axis) 
and a (2+1)-dimensional expansion along the longitudinal direction ($z$ axis)
as well as along the transverse  direction with cylindrical symmetry ($r$
axis).

To evaluate the conservation laws $\PD_\m N^\m=0$ and $\PD_\m T^{\m\n}=0$ in a
particular coordinate  system we calculate $N^\m$ and $T^{\m\n}$ components 
in that coordinate system and convert these components into physical
components using Eqs.~(\ref{eq:vcon}), (\ref{eq:vco}),  (\ref{eq:tco}) and
(\ref{eq:tcon}) from Appendix \ref{append:physcomp}.  Then we  apply 
Eqs.~(\ref{eq:cyltransf}) and (\ref{eq:sphtransf}) from Appendix
\ref{append:physcomp}.

For (1+1)-dimensional expansion along the $z$-axis in planar geometry  
derivatives with respect to $(x,y)$ must be identically zero by symmetry. So we
need to calculate only terms containing derivatives in $(t,z)$. 
Similarly  for (1+1)-dimensional spherical symmetric flow the terms in
$(\PD/\PD \q)$ and  $(\PD/\PD \f)$ vanish identically while for
(1+1)-dimensional cylindrical  symmetric flow the terms in $(\PD/\PD \f)$ and 
$(\PD/\PD z)$ also vanish identically. So we need to calculate only terms
containing $(\PD/\PD t)$ and $(\PD/\PD r)$.

\subsection{(1+1)-dimensional expansion in planar geometry}

In high energy nuclear collisions one often considers (1+1)-dimensional
expansion of the produced matter along the beam or collision axis taken to be
the $z$-axis. In this case there is only one non-vanishing spatial component
of  the 4-velocity. In relativistic ideal fluid dynamics it is sufficient to
decompose the net charge 4-current and the energy-momentum stress tensor using
only one 4-vector, namely the 4-velocity in addition to the metric tensor.
However in relativistic non-ideal fluid dynamics due to additional dissipative
fluxes that are orthogonal to the 4-velocity, the decomposition of the net
charge 4-current and the energy-momentum stress tensor requires additional
4-vectors in order for the decomposed quantities to be expressed in terms
of the 4-vectors, the metric tensor and the local scalar quantities. Since in
the local rest frame the components of the dissipative fluxes are spatial the
additional 4-vectors are space-like while the 4-velocity is time-like. In
(1+1)-dimensional expansion we need only one more 4-vector in addition to the
4-velocity. Note that the additional 4-vectors are not needed if one wants to
keep the dissipative fluxes in their lab frame tensorial form. Only if we want
to write them in their local rest frame scalar form in the energy-momentum stress
tensor will we need additional 4-vectors. This is possible for simple cases such
as (1+1)-dimensional expansions.

From the line element 
\be
d s^2 = dt^2 - dx^2 -dy^2 -dz^2
\ee
the Minkowski metric is $g^{\m\n}=\mbox{diag} (1,-1,-1,-1)$. 
For a local observer comoving with the fluid, the two 4-vectors are
\bea
\widehat{u}^{\m}(t,z,x,y) &=& (1,0,0,0) ~,\\
\widehat{m}^{\m}(t,z,x,y) &=& (0,1,0,0) ~,
\eea
in the local rest frame: one is the 4-velocity which is time-like,
$\widehat{u}_\m \widehat{u}^\m=1$, and the other one is space-like,
$\widehat{m}_\m \widehat{m}^\m = -1$, and it is orthogonal to the 4-velocity,
$\widehat{m}^\m \widehat{u}_\m =0$. Hence the local rest frame is defined by
$(\widehat{u}^\m = \delta^\m_t$, $\widehat{m}^\m = \delta^\m_x$). 
Note that in this simple (1+1)-dimensional case there is only one independent
component of the heat flux, taken to be the $z$ component. In the local rest
frame the heat flux is  $\widehat{q}^\m = (0,\cq^{z},0,0) = q\delta^\m_x =
q\widehat{m}^\m$ which we simply denote by $q$, the heat flow. In the local rest
frame the shear tensor takes the form 
\begin{equation}
\widehat{\pi }^{\mu \nu }=\left( 
\begin{array}{cccc}
0 & 0 & 0 & 0 \\ 
0 & \t^{zz} & 0 & 0 \\ 
0 & 0 & \t^{xx} & 0 \\ 
0 & 0 & 0 & \t^{yy}
\end{array}
\right) ,  \label{eq:pimunu-lrfp}
\end{equation}
where $\t^{\m\n}$ represent the shear stress tensor components in the local rest
frame of the fluid. 
There is also one independent component of the shear tensor in this simple
(1+1)-dimensional problem, taken to be the $zz$ component  $\widehat{\pi}^{zz} =
\t^{zz} =\pi$ which we simply denote it by $\pi$.  The tracelessness property of
the shear stress tensor then implies that in Eq.~(\ref{eq:pimunu-lrfp})
$\widehat{\pi}^{xx}=\t^{xx}=\widehat{\pi}^{yy} = \t^{yy} = -{\pi\over 2}$. Note also
that the bulk viscous pressure enters the non-ideal part of the energy-momentum
tensor as $-\mPi \btu^{\m\n}$ and in the local rest frame it is $-\mPi
\widehat{\btu}^{\m\n}$ where $\widehat{\btu}^{\m\n} = g^{\m\n}-\widehat{u}^\m
\widehat{u}^\n$.

The net charge 4-current in the local rest frame is
\be
\widehat{N}^\m = (n, 0, 0,0) ~.
\ee
The energy--momentum tensor in  the local rest frame is given by 
\be 
    \widehat{T}^{\m\n} = 
    \begin{pmatrix}  
    \eps        &\quad q   &\quad 0
                      &\quad 0 
\\
    q  &\quad \cp_\perp      &\quad 0  
                      &\quad 0  
\\
    0 &\quad 0    &\quad   \cp_\perp     
                      &\quad 0  
\\ 
    0 &\quad 0   &\quad 0 
                      &\quad \cp_z
    \end{pmatrix}  \label{eq:TmunuLRF1}\;.
\ee
where $\cp_z = p+\pi+\mPi$, $\cp_\perp = p-\pi/ 2 +\mPi$.  That is, the net
charge 4-current and the  local rest frame energy--momentum tensor can be
decomposed as
\bea
\widehat{N}^\m &=& n \widehat{u}^\m ~,\\
\widehat{T}^{\m\n} &=& \left(\eps+\cp_\perp\right)\widehat{u}^\m \widehat{u}^\n - \cp_\perp g^{\m\n}
+\left(\cp_z-\cp_\perp\right)\widehat{m}^{\m} \widehat{m}^\n + 2q\, \widehat{m}^{(\m}\widehat{u}^{\n)} ~,
\label{eq:Tmn-lrf}
\eea
The dynamics of the system can be studied by applying a Lorentz boost with
$v_z$ in the $z$ direction. Thus the net charge 4-current and the
energy-momentum stress tensor as measured by an
observer with velocity $v_z$ with respect to the fluid configuration are given
by
\bea
N^\m &=& n u^\m ~,\\
T^{\m\n} &=& (\eps+\cp_\perp) u^\m u^\n - \cp_\perp g^{\m\n} + (\cp_z-\cp_\perp)
m^\m m^\n +2 q \,m^{(\m}u^{\n)} ~,\label{eq:Tmn-calc}
\eea
where $u^\mu$ and $m^\m$ are given by 
\bea
u^\m(t,z,x,y) &=& (\g, \g v_z,0,0) \label{eq:4v1d}~,\\
m^\m(t,z,x,y) &=& (\g v_z,\g,0,0)\label{eq:4m1d}~,
\eea
and $v_z$ is the fluid 3--velocity in the $z$ direction and
$\g=\left(1-v_z^2\right)^{-1/2}$.  Note that $q^\m u_\m = 0$ and $q =
\sqrt{-q^\m q_\m}$. Note also that $m^\m = q^\m/q$. Explicit components of the
net charge 4-current and the energy-momentum tensor read 
\bea
N^0 &=& \g n ~,\label{eq:N0}\\
N^z &=& N^0 v_z~,\label{eq:Nz}\\
T^{00} &=& \cw \g^2 - \cp_z + 2 q\g^2 v_z ~,\label{eq:T001}\\
T^{0z} &=& \cw \g^2 v_z  +  q\g^2(1+ v_z^2) ~,\label{eq:T0z1}\\
T^{zz} &=& \cw\g^2 v_z^2 + \cp_z +2 q \g^2 v_z~, \\
T^{xx} &=& T^{yy} = \cp_\perp ~,
\eea 
with $\cw\equiv \eps+\cp_z$. 
Using Eqs.~(\ref{eq:N0}), (\ref{eq:T001}) and (\ref{eq:T0z1}) 
the local velocity, energy density and net charge density can be obtained from
\be
N^0 = \g n \sks \eps = T^{00} - (T^{0z}+q)v_z \sks 
T^{0z} = (T^{00}+\cp_z)v_z + q
~.
\ee
That is,
\bea
v_z &=& {T^{0z} - q \over T^{00}+\cp_z} ~,\\
\eps &=& T^{00} - {{T^{0z}}^2 -q^2 \over T^{00}+\cp_z} ~,\\
n &=& N^0 \sqrt{1-v_z^2} ~.
\eea	   
The net charge and the energy--momentum conservation equations can be written as  
\bea
\PD_\m N^\m \equiv 0 &\Lra& \PD_t N^0 +\PD_z (N^0 v_z) = 0 ~.\\
\PD_\m T^{\m 0}\equiv0 &\Lra& {\PD \over \PD t} T^{00} 
                   +{\PD \over \PD z} \left\{\left(T^{00}+\cp_z
		   \right)v_z+q\right\}=0 ~,\\
\PD_\m T^{\m z}\equiv0 &\Lra& {\PD \over \PD t} T^{0z} 
                   +{\PD \over \PD z} \left\{\left(T^{0z}+q
		   \right)v_z+\cp_z\right\}=0 ~,
\eea	
The Maxwell-Cataneo transport equations in simple one--dimensional expansion
takes the form
\bea
\t_\mPi\dot{\mPi} +\mPi &=& -\z\q ~,\\
\t_q\dot{q} + q &=& -\k T\left({T^{\pr}\over T}+a\right) ~,\\
\t_\pi\dot{\pi} +\pi &=& -2\cdot2\h\s ~, \label{eq:piT}
\eea
where
\bea
\dot{f} &\equiv& \left[\g{\PD \over \PD t}+\g v_z {\PD \over \PD
z}\right] f~,\\
f^{\pr} &\equiv& \g\left[\g v_z{\PD \over \PD t}+\g {\PD \over \PD
z}\right] f~,\\
\q &\equiv& {\PD \g \over \PD t} + {\PD \g v_z \over \PD z} ~,\\
\s &=& \s^{xx} = {1\over 3}\q ~.
\eea

The calculational frame components of the shear tensor $\pi^{\m\n}$ are related
to the local rest frame components $\t^{\m\n}$ as follows
\bea
\pi^{zz} &=& \g^2 \t^{zz} = \g^2 \pi ~, \label{eq:longboost}\\
\pi^{xx} &=& \pi^{yy} = \t^{xx}=\t^{yy} = -{\p\over 2} ~,\label{eq:transboost}
\eea
which can be read directly from Eqs~(\ref{eq:Tmn-lrf}) and (\ref{eq:Tmn-calc}).
Thus due to its simplistic structure, we numerically evolve the transverse
components  ($\pi^{xx}$ or $\pi^{yy}$) since they do not change under the
boost and thus they do not involve the Lorentz gamma factors. Thus we
numerically solve  Eq.~(\ref{eq:piT}). Then we can obtain the longitudinal
component $\pi^{zz}$ from which the local rest frame quantity $\t^{zz}$ follows
from Eq.~(\ref{eq:longboost}). The local rest frame quantity $\t^{zz}$ can also
be obtained from the tracelessness of $\pi^{\m\n}$ which implies that
\be
\t^{zz} = -(\t^{xx}+\t^{yy}) ~.
\ee
\subsection{ (2+1)-dimensional axisymmetric expansions}
(2+1)-dimensional axisymmetric expansions are (2+1)-dimensional in Cartesian
coordinates or planar geometry (the flow 4-vectors such as the 4-velocity are
functions of time and two spatial coordinates) but they are (1+1)-dimensional
in one dimensional cylindrically/spherically symmetric expansions. In this section
we consider systems that exhibit cylindrical symmetry though the results can be
generalized to include  spherical (fireball) symmetry for radial expansions. In
cylindrical coordinates all derivatives with respect to the axial and
azimuthal  directions are zero, and the 4-vectors are functions of time  and
radial coordinates only.  We consider the case where the azimuthal and axial
components of the 4-flow vectors are zero and the only spatial component is the
radial component.

Note that since this is effectively a (1+1)-dimensional problem as a result of
symmetry considerations there is only one independent component of heat flux
taken to be the radial component  which in the local rest frame is
$\widehat{q}^\m = (0,\cq^r\E_r,0)=(0,q\E_r,0)$ and there are two independent
components of the shear stress tensor taken to be the $zz$ and $\f\f$ 
components  in cylindrical polar coordinates. Here $\E_r=(\cos\f,~\sin\f)$.  We
will derive the equations for the case with cylindrical symmetry, and then restate important
results so that they also apply to spherical symmetry and Cartesian
coordinates.

Starting from the planar geometry (where the 4-velocity has two non-vanishing
spatial components, namely the $x$ and $y$ component) the local rest frame 
shear stress tensor is transformed into the cylindrical coordinate form 
\begin{equation}
\widehat{\pi }^{\m\n} =\left( 
\begin{array}{cccc}
0 &    0   & 0      &   0 \\ 
0 & \t^{xx} & \t^{xy} &   0 \\ 
0 & \t^{yx} & \t^{yy} &   0 \\ 
0 &    0   & 0      & \t^{zz}
\end{array}
\right) =
\left( 
\begin{array}{cccc}
0 & 0 & 0 & 0 \\ 
0 & [\t^{rr}-\t^{\f\f}]\cos^2\f+\t^{\f\f} &[\t^{rr}-\t^{\f\f}]\cos\f\sin\f   & 0\\ 
0 & [\t^{rr}-\t^{\f\f}]\cos\f\sin\f  & [\t^{rr}-\t^{\f\f}]\sin^2\f+\t^{\f\f} & 0\\ 
0 & 0 & 0 & \t^{zz}
\end{array}
\right) 
\label{eq:pimunu-lrf}
\end{equation}
The local rest frame net charge 4-current and energy-momentum stress tensor
takes the form
\bea
\widehat{N}^\m &=& (n,0,0,0)~,\\
    \widehat{T}^{\m\n} &=& 
    \begin{pmatrix}  
    \eps   &\quad \cq^x     &\quad \cq^y    &\quad 0 \\
     \cq^x &\quad \cp_{xx}  &\quad \cp_{xy}  &\quad 0 \\
    \cq^y  &\quad \cp_{yx}  &\quad \cp_{yy}  &\quad 0 \\ 
    0  &\quad     0  &\quad     0  &\quad \cp_{zz}
    \end{pmatrix}  \\
&=&
    \begin{pmatrix}  
    \eps   &\quad \cq^r\cos\f     &\quad \cq^r\sin\f     &\quad 0 \\
     \cq^r\cos\f &\quad [\cp_r-\cp_\f]\cos^2\f+\cp_\f  &\quad
     [\cp_r-\cp_\f]\cos\f\sin\f  &\quad 0  \\
    \cq^r\sin\f  &\quad [\cp_r-\cp_\f]\cos\f\sin\f   &\quad
    [\cp_r-\cp_\f]\sin^2\f+\cp_\f  &\quad 0 \\ 
    0  &\quad  0 &\quad  0 &\quad \cp_z
    \end{pmatrix}  
\label{eq:TmunuLRF2} ~, 
\eea
where $\cp_r = p+\mPi+\t^{rr}$, $\cp_\f = p+\mPi+\t^{\f\f}$, $\cp_z =
p+\mPi+\t^{zz}$ and $\cp_{ij} = p+\mPi+\t_{ij}$. The pressure tensor is defined
such that $\cp^{ij}$ is the net rate of transport, per unit area of a surface
oriented perpendicular to the $j$th coordinate axis, of the $i$th component of
momentum.

Since in this $(t,x,y,z)$ coordinate system we have two non-vanishing spatial
components of 4-velocity 
there are now more vectors required to construct a complete set of tensors with
respect to which $T^{\m\n}$ should be decomposed.  The set of four-vector
fields that we need are  $u^\m$, $l^\m$ and $m^\m$ which in the local rest
frame takes the form 
\bea
\widehat{u}^\m (t,x,y,z) &=& (1, 0,0,0) ~,\\ 
\widehat{l}^\m (t,x,y,z) &=& (0,0,0,1) ~,\\
\widehat{m}^\m (t,x,y,z) &=& (0,\E_r,0) ~.
\eea
with the properties $\widehat{u}_\m \widehat{u}^\m = 1$, $\widehat{l}_\m
\widehat{l}^\m = \widehat{m}^\m \widehat{m}_\m =-1$ and $\widehat{u}_\m
\widehat{l}^\m =\widehat{u}_\m \widehat{m}^\m = \widehat{l}_\m \widehat{m}^\m
=0$. One vector is time like and the other two are space-like. The space-like
vectors are all orthogonal to each other and to the time-like vector. The two
additional space-like 4-vectors are now required since we have flows in two
spatial directions. Also unlike in the ideal fluid dynamics we now have the
coupling of the two non-vanishing velocity components as a result of dissipative
effects. 
Thus in the local rest frame the net charge 4-current and the energy momentum
tensor can be decomposed as
\bea
\widehat{N}^\m &=& n \widehat{u}^\m ~\\
\widehat{T}^{\m\n} &=& (\eps+\cp_\f) \widehat{u}^\m \widehat{u}^\n +
\cp_\f g^{\m\n} + (\cp_z-\cp_\f) \widehat{l}^\m \widehat{l}^\n  
+ (\cp_r-\cp_\f)\widehat{m}^\m \widehat{m}^\n + 2\cq^r \widehat{u}^{(\m} \widehat{m}^{\n)} ~.
\label{eq:Tmunu-lrf}
\eea

Relativistic dynamics of the viscous heat-conducting fluid is accomplished by
a Lorentz boost in the radial direction. For an observer moving with velocity
$v_r$ in the radial direction with respect to the fluid configuration the net
charge 4-current and the energy momentum as measured by such an observer are
\bea
N^\m &=& n u^\m ~,\\
T^{\m\n} &=& (\eps+\cp_\f) u^\m u^\n + \cp_\f g^{\m\n} 
+ (\cp_z-\cp_\f) l^\m l^\n  + (\cp_r-\cp_\f) m^\m m^\n + 2\cq^r u^{(\m} m^{\n)} ~.
\label{eq:Tmunu-calc}
\eea
where
\bea 
u^\m (t,x,y,z) &=& \left(\g, \g v_r \E_r\right) ~, \label{eq:4v-calc} \\
l^\m (t,x,y,z) &=& \left(\g v_r, 0, 0, \g \right) ~, \label{eq:4l-calc} \\
m^\m (t,x,y,z) &=& \left(\g v_r, \g \E_r\right) ~. \label{eq:4m-calc} 
\eea
The relationship between the local rest frame and calculational frame
components of the net charge 4-current and the energy-momentum stress tensor
can be read off from Eqs.~(\ref{eq:Tmunu-lrf}) and (\ref{eq:Tmunu-calc}).  The
components of the energy--momentum tensor in the projected coordinate  system
are 
\bea
T^{00} &=& \cw \g^2 - \cp_r + 2 \cq^r \g^2 v_r~,\\
T^{0r} &=& \cw \g^2 v_r  +  \cq^r \g^2(1+ v_r^2)~,\\
T^{rr} &=& \cw\g^2 v_r^2 +\cp_r +2 \cq^r \g^2 v_r ~,\\
T^{\f\f} &=& \cp_\f ~, \\
T^{zz} &=& \cp_z ~,
\eea
with $\cw \equiv \eps+\cp_r$ and the usual components are obtained in terms of
these as
\bea
T^{xx} &=& T^{rr} \cos^2 \f +T^{\f\f} \sin^2 \f ~\\
T^{yy} &=& T^{rr} \sin^2 \f +T^{\f\f} \cos^2 \f ~\\
T^{xy} &=& [T^{rr}-T^{\f\f}] \cos \f \sin \f ~,\\
T^{0x} &=& T^{0r}\cos \f \sks T^{0y} = T^{0r}\sin \f~,\\
T^{xz} &=& T^{rz}\cos \f \sks T^{yz} = T^{rz}\sin \f~.
\eea
The components of 4-vectors and 2nd-rank energy-momentum stress tensor
presented here are physical components, and are identical to the components
measured with respect to an orthonormal tetrad in a curvilinear, in this case
cylindrical, coordinate system. The contravariant or covariant components can be
obtained from the transformation rules given in Appendix \ref{append:physcomp}
starting from cylindrical coordinates.

The local velocity, energy density and net charge density can be obtained from
\be
\eps = T^{00} - (T^{0r}+\cq^r)v_r \sks T^{0r} = (T^{00}+\cp_r)v_r + \cq^r \sks
N^0 = \g n ~,
\ee
that is,
\bea
v_r &=& {T^{0r} - \cq^r \over T^{00}+\cp_r} ~,\\
\eps &=& T^{00} - {{T^{0r}}^2 -{\cq^r}^2 \over T^{00}+\cp_r} \enspace~,\\
n &=& (1-v_r^2)^{1/2} N^0~.
\eea	   
For the space--time derivatives we use the Eq.~(\ref{eq:cyltransf}).
The net charge conservation $\PD_\m N^\m =0$ and 
the energy--momentum conservation $\PD_\m T^{\m\n} = 0$ can be written as
\bea
\PD_\m N^\m \equiv 0 &\Lra& \PD_t N^0 + {1\over r^\a} \PD_r (r^\a N^0 v_r) = 0 ~,\\
\PD_\m T^{\m 0} \equiv 0 &\Lra& \PD_t T^{00} 
+ {1\over r^\a} \PD_r \{r^\a T^{r0}\} = 0 ~,\\
\PD_\m T^{\m r} \equiv 0 &\Lra& \PD_t T^{0r} + {1\over r^\a}\PD_r \{r^\a T^{rr}\} 
-{\alpha\over r} T^{\f\f} =0 ~,
\eea
where $\alpha = 1,2$ for cylindrical and spherical geometry respectively. 
These equations can be simplified and written in a better way, suitable for 
numerical purposes, as follows
\bea
&&{\PD \over \PD t} N^0 + {\PD \over \PD r}( N^0 v_r) = -{\a\over r} N^0 v_r ~,\\
&&{\PD \over \PD t} T^{00} 
                   +{\PD \over \PD r} \left\{\left(T^{00}+\cp_r
		   \right)v_r+q\right\}
		   = - {\a\over r} \left(T^{00}+\cp_r\right)v_r -{\a\over r} q~,\\
&&{\PD \over \PD t} T^{0r} 
                   +{\PD \over \PD r} \left\{\left(T^{0r}+q
		   \right)v_r+\cp_r\right\}
		   = -{\a\over r}\left(T^{0r}+q\right)v_r 
		   -{\a\over r} \left(\cp_r-\cp_\f\right) ~.
\eea	
Note that the structure of these equations is similar to pure
(1+1)-dimensional expansion except for the geometric terms on the right hand
side. These terms results from the transformation using Eqs.~(\ref{eq:cyltransf}) and 
(\ref{eq:sphtransf}).

For the transport equations recall that there is only one independent
component  of the heat flux and two of the shear stress tensor. For the heat 
flux this translates into the equation for $q$, the heat flow. For the shear
stress tensor we choose the $zz$ and the $\f\f$ components. This is just for
convenience since these choices makes the calculations more tractable. The
simplicity comes from the property that these transverse components of the
shear stress tensor do not change under the boost and hence they do not pick up
the Lorentz gamma factors which would otherwise render the solution more
difficult. In spite of the fact that not all of the equations are independent
we shall present them all, since depending on the problem under consideration,
it may be more advantageous using one set instead of the other. The weak
coupling limit of the causal transport
equations with the cylindrical symmetry takes the form
\bea
\t_\mPi\dot{\mPi} +\mPi &=& -\z\mTheta  ~,\\
\t_q\dot{q^r} + q^r &=& -\k T\left({T^\pr\over T} + a\right)  ~,\\
\t_\pi\dot{\pi}^{rr} +\pi^{rr} &=& -2\h \s^{rr}  ~,\\
\t_\pi\dot{\pi}^{\f\f} +\pi^{\f\f} &=& -2\h \s^{\f\f}  ~, \\
\t_\pi\dot{\pi}^{zz} +\pi^{zz} &=& -2\h \s^{zz}  ~,
\eea
where
\bea
\mTheta &\equiv& \theta + \alpha {\g v_r\over r}~,\\
\theta &\equiv&  {\PD \over \PD t}\gamma +{\ppr} \gamma v_r ~,\\
\dot{f} &\equiv& \left[\g{\PD \over \PD t}+\g v_r {\PD \over \PD
r}\right] f~,\\
f^{\pr} &\equiv& \g\left[\g v_r{\PD \over \PD t}+\g {\PD \over \PD
r}\right] f~,\\
\s^{rr} &=& -\g^2 \left[-\a {\g v_r\over r} +{2\over 3}\mTheta \right]~,\\
\s^{\f\f} &=& -\a {\g v_r \over r} + {1\over 3} \mTheta ~,\\
\s^{zz} &=& {1\over 3} \mTheta ~.
\eea
The relationship between the  calculational frame and the local rest frame
components of the shear tensor can be read off from Eqs.~(\ref{eq:Tmunu-calc})
and
(\ref{eq:Tmunu-lrf}). In particular
\bea
q^r &=& \g \cq^r =\g q~,\\
\pi^{rr} &=& \g^2 \t^{rr}~,\\
\pi^{zz} &=& \t^{zz}  ~,\\
\pi^{\f\f} &=& \t^{\f\f}~.
\eea
The orthogonality and tracelessness of the shear stress tensor implies the
following relations respectively 
\bea
&&\pi^{rr}/\g_r^2+\pi^{\f\f}+\pi^{zz} = 0 ~,\\
&&\t^{rr}+\t^{\f\f}+\t^{zz} = 0~.
\eea
Including the bulk equation we only have three transport equations.  Again
these equations have the same structure as in (1+1)-dimensional case, except
for geometrical terms. Note that due to symmetry in the problem it can be shown
that in Eqs.~(\ref{eq:bulk})-(\ref{eq:shear})
\bea
\t_q\omega^{r\n} q_\n &=& \t_q\Delta^{r\a}\Delta^{\n\b} \PD_{[\b}u_{\a]} =
\t_q\Delta^{r\a} \PD_{[\b}u_{\a]} q^\b = 0 \enspace~,\\
2\eta\pi^{\a(\m}\omega^{\n)}_\a &=& 0 ~.
\eea
This is also the case with spherically symmetric expansions. 
One can deduce that in (1+1)-dimensional cylindrically symmetric flow it is not
possible to choose a scalar viscous pressure as is the case in (1+1) planar
geometry and (1+1)-dimensional spherically symmetric flows. This is because
viscous effects originate from a tensor, which is not, in general, isotropic
even for (1+1)-dimensional flows.

In going from the Cartesian coordinate system to the projected coordinate
system one notices that this is similar to having started with the pure  
$(t,r,\f,z)$ cylindrical coordinate system in which the 
dynamics of the system are obtained by applying a Lorentz boost in
the radial direction with
\bea
u^\m (t,r,\f,z) &=& (\g, \g v_r,0,0) ~,\\
m^\m (t,r,\f,z) &=& (\g v_r, \g,0,0) ~, 
\eea
as done in Appendix \ref{append:sphsym} for the spherical symmetry case using
the mixed tensor $T^\m_\n$. 
Note that since in this case we have only one non-vanishing spatial component
of 4-velocity, we now need only one additional spatial 4-vector for
energy-momentum stress tensor decomposition. 

\subsection{(3+1)--dimensional axisymmetric expansion: Firebarrel expansion}

We now consider the relativistic dynamics of non-ideal fluids in firebarell
geometry. (3+1)-dimensional axisymmetric flows are (3+1)--dimensional with
respect to Cartesian coordinates i.e., the velocity components are functions of
time and  all three spatial coordinates, but they are only (2+1)--dimensional
in cylindrical coordinate system. All derivatives with respect to azimuthal
direction are zero, and all 4-vector components such as the 4-velocity
components are functions of only the time, axial and radial coordinates, 
$(t,z,r)$. In case without swirl, the azimuthal velocity component is zero
everywhere. As it is much easier to work with (2+1) independent variables than
(3+1), for axisymmetric flows, it makes sense to work in a cylindrical
coordinate rather than Cartesian one.

Let us consider a system undergoing expansion along the longitudinal direction
taken to be the $z$-axis and at the same time expanding along the transverse 
(radial) direction taken to be the $r$-axis. The 4-velocity in cylindrical polar
coordinates takes the form 
\be
u^\m (t,x,y,z) = (\g, \g v_r\cos \f,\g v_r\sin \f,\g v_z) ~, 
\ee
with 
\bea
\gamma &=& (1-v^2)^{-1/2} ~,\\
v^2 &=& v_z^2 + v_r^2 ~.
\eea

The net charge density, energy density and momentum density components read  
\bea
N^0 &=& \g n ~,\\
T^{00} &=& (\eps+\cp)\g^2 -\cp +2\g (q^z v_z +q^r v_r) + \pi^{rr}+\pi^{\f\f}
+\pi^{zz} ~,\\
T^{0z} &=& (\eps+\cp)\g^2 v_z  +\g q^z(1+ v_z^2) + q^r \g v_r v_z +
\pi^{zz} v_z + \pi^{zr}v_r ~,\\
T^{0r} &=& (\eps+\cp)\g^2 v_r  +\g q^r(1+ v_r^2) +q^z \g v_z v_r +
(\pi^{rr} + \pi^{\f\f}) v_r + \pi^{rz} v_z ~,
\eea
while the other components needed in the conservation equations can be
written in term of the above equations
\bea
N^z &=& N^0 v_z \sks N^r = N^0 v_r ~,\\
T^{zz} &=& (\eps+\cp)\g^2 v_z^2 +\cp + 2 q^z \g v_z + \pi^{zz} ~,\\
T^{rr} &=& (\eps+\cp)\g^2 v_r^2 +\cp + 2 q^r \g v_r + \pi^{rr} ~,\\
T^{rz} &=& (\eps+\cp)\g^2 v_r v_z  +  \g (q^z v_r +q^r v_z) + \pi^{rz} ~,\\
T^{\f\f} &=& \cp +\pi^{\f\f} ~,\\
T^{xx} &=& [T^{rr}-T^{\f\f}]\cos^2 \f + T^{\f\f} ~\\
T^{yy} &=& [T^{rr}-T^{\f\f}]\sin^2 \f + T^{\f\f}  ~\\
T^{xy} &=& [T^{rr}-T^{\f\f}] \cos \f \sin \f ~,\\
T^{0x} &=& T^{0r}\cos \f \sks T^{0y} = T^{0r}\sin \f~,\\
T^{xz} &=& T^{rz}\cos \f \sks T^{yz} = T^{rz}\sin \f~.
\eea
Note that we have replaced $\pi^{00}$ and $\pi^{0i}$ by their spatial
components using the orthogonality and the tracelessness of the shear tensor
$\pi^{\m\n}$. In particular
\bea
\pi^{00} &=& \pi^{xx} +\pi^{yy} + \pi^{zz} = \pi^{rr} + \pi^{\f\f} + \pi^{zz} ~,\\
\pi^{0z} &=& \pi^{zz} v_z + \pi^{zr} v_r ~,\\
\pi^{0r} &=& [\pi^{rr} + \pi^{\f\f}] v_r +\pi^{rz} v_z ~,
\eea
where the first equation for $\pi^{00}$ comes from the tracelessness of
$\pi^{\m\n}$ ($\pi^\n_\n =0$) and the others comes from the orthogonality
condition ($\pi^{\m\n} u_\n =0)$. Note also that in the projected coordinate
system one could replace $\pi^{00}$ by the spatial components using the
orthogonality relations giving
\be
\pi^{00} = \pi^{rr} v_r^2 + 2 \pi^{rz} v_r v_z +\pi^{zz} v_z^2 ~,
\ee
from which the tracelessness conditions now becomes
\be
{\pi^{rr}+\pi^{\f\f} \over \g_r^2} -2\pi^{rz} v_r v_z + { \pi^{zz}\over \g_z^2}
 =0~.
\ee

For simplicity and in the rest of this section we will now consider the
dynamics of a relativistic viscous non-heat-conducting system in this
firebarrel geometry.  For the numerical purposes we rewrite the momentum
components of the energy-momentum stress tensor in terms of the energy
component and the spatial components in terms of the momentum components
\bea
T^{0z} &=& (T^{00} +\cp -[\pi^{rr} +\pi^{\f\f}]) v_z +\pi^{zr}v_r
~,\label{eq:vT0z}\\
T^{0r} &=& (T^{00} +\cp -\pi^{zz}) v_r +\pi^{rz} v_z ~,\label{eq:vT0r}\\
T^{rr} &=&  T^{0r} v_r +\cp +\pi^{rr}/\g_r^2 -\pi^{\f\f} v_r^2 - \pi^{rz} v_r v_z ~,\\
T^{zz} &=&  T^{0z} v_z +\cp +\pi^{zz}/\g_z^2 - \pi^{zr} v_z v_r ~,\\
T^{zr} &=&  T^{0z} v_r + \pi^{zr}/\g_r^2  +\pi^{zz} v_z v_r ~,\\
T^{rz} &=&  T^{0r} v_z + \pi^{rz}/\g_z^2 - [\pi^{rr}+\pi^{\f\f}] v_r v_z
\eea
In differential form, the (2+1)--dimensional conservation equations for net
charge  density, energy density and momentum density written in cylindrical
coordinate system read
\bea
\PD_\m N^\m  &\equiv& \PD_t N^0 + \PD_z N^z +{1\over r}\PD_r\{r N^r\}  ~,\\
\PD_\m T^{\m 0}&\equiv& \PD_t T^{00} + \PD_z T^{0z} + {1\over r} \PD_r\{ r T^{0r}\} = 0 ~,\\
\PD_\m T^{\m z}&\equiv& \PD_t T^{z0} + \PD_z T^{zz} + {1\over r} \PD_r\{ r T^{zr}\} = 0 ~,\\
\PD_\m T^{\m r}&\equiv& \PD_t T^{r0} + \PD_z T^{rz} + {1\over r} \PD_r\{ r T^{rr}\} -{1\over r}
T^{\phi\phi} = 0 ~.
\eea
For numerical purposes we write the above set of conservation equations in
terms of the conserved quantities in the calculational frame, namely the net
charge density, the energy density and the momentum densities 
\bea
\PD_t N^0 &=& -\PD_z (N^0 v_z) -{1\over r}\PD_r\{r N^0 v_r\} = 0 ~,\\
\PD_t T^{00} &=& -\PD_z\{T^{00}v_{sz}\} - \PD_z\{T^{00}v_{sr}\} -{1\over
	     r}T^{r0} ~,\\
\PD_t T^{z0} &=& -\PD_z\{(T^{z0} v_z + \cp + \pi^{zz}/\g_z^2 - \pi^{zr} v_z v_r\}
                 -\PD_r\{ T^{z0} v_r + \pi^{zr}\g_r^2 +\pi^{zz} v_z v_r \} \nonumber\\
		 &&-{1\over r} \{T^{0z} v_r + \pi^{zr}/\g_r^2 + \pi^{zz} v_z v_r\}~,\\
\PD_t T^{r0} &=& -\PD_z\{T^{r0} v_z + \pi^{rz}/\g_z^2 
                                    - [\pi^{rr}+\pi^{\f\f}]v_r v_z\}
                 -\PD_r\{(T^{r0}v_r + \cp + \pi^{rr}/\g_r^2 - \pi^{\f\f} v_r^2
		 -\pi^{rz} v_r v_z\}\nonumber\\
		 &&-{1\over r} \{(T^{r0}v_r +\cp 
		 + [\pi^{rr}+\pi^{\f\f}]/\g_r^2 -\pi^{rz} v_r v_z\}~,
\eea
where 
\be
v_{sz} = {T^{0z}\over T^{00}} \sks v_{sr} = {T^{0r}\over T^{00}} ~. 
\ee
The local velocities can be obtained from simultaneously solving
Eqs.~(\ref{eq:vT0z}) and
(\ref{eq:vT0r}) while the local net charge density and energy density are
obtained from
\bea
n &=& \cn (1- v^2)^{1/2} ~,\\
\eps &=& T^{00} - T^{0z} v_z - T^{0r} v_r ~, 
\eea
For the transport equations we need the bulk pressure equation and the
equations for the non-vanishing shear stress tensor components that are needed
in the conservation equations
\bea
\PD_t \mPi + [v_z\PD_z + v_r\PD_r]\mPi &=& -{1\over \g \t_\Pi}
\left(\zeta\mTheta - \mPi \right) ~,\\
\PD_t \pi^{rr} + [v_z\PD_z + v_r\PD_r]\pi^{rr} &=& {1\over \g \t_\pi}
\left(2\eta\s^{rr} - \pi^{rr} \right) ~,\\
\PD_t \pi^{zz} + [v_z\PD_z + v_r\PD_r]\pi^{zz} &=& {1\over \g \t_\pi}
\left(2\eta\s^{zz} - \pi^{zz} \right) ~,\\
\PD_t \pi^{rz} + [v_z\PD_z + v_r\PD_r]\pi^{zr} &=& {1\over \g \t_\pi}
\left(2\eta\s^{zr} - \pi^{zr} \right) ~,\\
\PD_t \pi^{\f\f} + [v_z\PD_z + v_r\PD_r]\pi^{\f\f} &=& {1\over \g \t_\pi}
\left(2\eta\s^{\f\f} - \pi^{\f\f} \right) ~,
\eea
where 
\bea
\mTheta  &=& \PD_t\g + \PD_z (\g v_z) + \PD_r (\g v_r) +{\g v_r\over r} ~,\\
\s^{rr} &=& \btd^r(\g v_r) + {1\over 3} (2+\g^2 v_r^2) \mTheta ~,\\
\s^{zz} &=& \btd^z (\g v_z) + {1\over 3} (1+\g^2 v_z^2)\mTheta ~,\\
\s^{rz} &=& {1\over 2} \left[\btd^r (\g v_z) + \btd^z (\g v_r) \right] +{1\over 3}
\g^2 v_r v_z \mTheta ~,\\
\s^{\f\f} &=& -{\g v_r \over r} +{1\over 3} \mTheta~,
\eea
with
\bea
\btd_z &=& \left[\PD_z +\g^2 v_z(\PD_t + v_z \PD_z + v_r \PD_r)\right] =-\btd^z~,\\
\btd_r &=& \left[\PD_r +\g^2 v_r(\PD_t + v_z \PD_z + v_r \PD_r)\right] =-\btd^r~.\\
\eea
Note that
\bea
\s^{xx} &=& \s^{rr} \cos^2 \f +\s^{\f\f}\sin^2 \f ~,\\
\s^{yy} &=& \s^{rr} \sin^2 \f + \s^{\f\f}\cos^2 \f ~,\\
\btd^x u^x + \btd^y u^y &=& \btd^r \g v_r - {\g v_r\over r}~.
\eea
The local rest frame dissipative fluxes are obtained from the above evolved
quantities
\be
\pi^{\m\n} = \L^\m_\a \L^\n_\b \widehat{\pi}^{\a\b} 
= \L^\m_\a \L^\n_\b \t^{\a\b} ~.
\ee
Note that due to symmetry considerations in our problem we only need to solve
the bulk pressure equation and the three independent components of the shear
stress tensor. 
In the non-relativistic limit the equations for mass density, energy and
momentum  conservation are
\bea
&&\PD_t \rho + \PD_z (\rho v_z) +{1\over r}\PD_r(r \rho v_r) = 0~,
\label{eq:rho}\\
&&\PD_t E + \PD_z\{(E+\cp_z) v_z +\pi^{zr} v_r\} 
+{1\over r}\PD_r\{r[(E+\cp_r)v_r + \pi^{rz} v_z]\} = 0~,\label{eq:ener}\\
&&\PD_t \cm^z + \PD_z\{\cm^z v_z+\cp_z\} +{1\over r} \PD_r\{ r(\cm^z v_r
+\pi^{zr})\} =0~,\label{eq:momz}\\
&&\PD_t \cm^r +\PD_z\{\cm^r v_z + \pi^{rz}\} 
+ {1\over r}\PD_r\{r[\cm^r v_r +\cp_r]\} - {1\over r}\cp_\f =0~,\label{eq:momr}
\eea
where
\bea
E &=& \epsilon + {1\over 2}\rho(v_z^2+v_r^2)~,\\
\epsilon &=& {1\over \Gamma -1}p~,\\ 
\cm^z &=& \rho v_z~,\\
\cm^r &=& \rho v_r~,\\
\cp_r &=& p+\mPi + \pi^{rr}~,\\
\cp_z &=& p+\mPi + \pi^{zz}~,\\
\cp_\f &=& p+\mPi + \pi^{\f\f}~.
\eea

The weak coupling limit of the causal transport equations in non-relativistic
limit takes the form
\bea
\t_\Pi \dot{\mPi} + \mPi = -\zeta\tnabla\cdot \tv ~,\\
\t_q    \dot{\tq} + \tq = -\lambda \tnabla T ~,\\
\t_\pi \dot{\tpi} + \tpi = -2\eta \tsigma ~.
\eea 
where here the  upper dot stands for the material or Lagrangian time derivative
(i.e., ${\dd / \dd t} = \partial/\partial t + \tv \cdot \tnabla$).
The explicit transport equations for the nonvanishing shear tensor components,
in the weak coupling limit are
\bea
\PD_t \pi^{zz} + [v_z\PD_z+v_r\PD_r]\pi^{zz} 
&=& -{1\over \t_\pi}(2\h \s^{zz}+\pi^{zz})~,\\
\PD_t \pi^{rr} + [v_z\PD_z+v_r\PD_r]\pi^{rr} 
&=& -{1\over \t_\pi}(2\h
\s^{rr}+\pi^{rr})~,\\
\PD_t \pi^{\f\f} + [v_z\PD_z+v_r\PD_r]\pi^{\f\f} 
&=& -{1\over \t_\pi}(2\h \s^{\f\f}+\pi^{\f\f})~,\\
\PD_t \pi^{rz} +[v_z\PD_z+v_r\PD_r]\pi^{rz} 
&=& -{1\over \t_\pi}(2\h \s^{rz}+\pi^{rz})~,
\eea
with
\bea
\s^{zz} &=& \PD_z v_z - {1\over 3}\tnabla\cdot \tv~,\\
\s^{rr} &=& \PD_r v_r - {1\over 3}\tnabla\cdot \tv~,\\
\s^{\f\f} &=& -{v_r\over r} - {1\over 3}\tnabla\cdot \tv~,\\
\s^{zr} &=& {1\over 2}\left(\PD_r v_z + \PD_z v_r\right) ~.
\eea

For numerical purposes we write the conservation equations,
Eqs.~(\ref{eq:rho})-(\ref{eq:momr}), as 
\bea
&&\PD_t \cn + \PD_z (\cn v_z) + \PD_r(\cn v_r) = -{\cn\over r}~, \\
&&\PD_t T^{00} + \PD_z\{T^{00} v_{sz}\} + \PD_r\{T^{00} v_{sr}\} 
= -{T^{0r}\over r}~,\\
&&\PD_t T^{0z} + \PD_z\{T^{0z} v_z + \cp_z\} + \PD_r\{ T^{0z} v_r+\pi^{zr}\} 
= -{(T^{0z} v_r -\pi^{zr})\over r}~,\\
&&\PD_t T^{0r} +\PD_z\{T^{0r} v_z +\pi^{rz}\} +\PD_r\{T^{0r} v_r + \cp_r\} 
= -{[T^{0r} v_r + \cp_r - \cp_\f]\over r} ~,
\eea
where
\bea
\cn &=& \rho \sks T^{00} = E \sks T^{0z} = \cm^z\sks T^{0r} = \cm^r ~,\\
v_{sz} &=& {T^{0z}\over T^{00}} \sks v_z = {T^{0z}\over \cn}~,\\
v_{sr} &=& {T^{0r}\over T^{00}}\sks v_r = {T^{0r}\over \cn}~.
\eea
In formulating the relativistic equations in this section we decided  to keep
the lab-frame quantities for the dissipative fluxes. Sometimes the local rest
frame equations are more complicated than the lab frame equations. The local
rest frame in this case of an effective (2+1)-dimensional flow contains Lorentz
contraction/dilation factors and higher powers of velocity which makes the
calculations and computations more difficult. The
transport/relaxation equations are readily written in their lab frame and if
one wants the corresponding local rest frame quantities one makes the necessary
Lorentz transformation.  In the non-relativistic limit the equations are cast
in laboratory frame quantities.
%
%
\section{Numerical aspects of relativistic nonideal fluid dynamics }
\label{sec:3d-code}
The equations of relativistic non-ideal fluid dynamics are truly formidable;
indeed they have never been solved for nontrivial problems except in simplified
cases such as scaling solutions \cite{AM2}.  Analytic solutions of relativistic
fluid dynamics are rare and in most cases of interest we must use numerical
methods \cite{MR}.

The complicated structure of relativistic dissipative fluid dynamics equations
makes it unattractive to use in modeling relativistic nuclear collisions.
Unlike in non-relativistic dissipative fluid dynamics we have time derivatives
appearing in more than one places. This makes computation more expensive.
Otherwise the system of equations given here are readily solved by existing
numerical schemes that are used to solve fluid dynamics problems. Knowledge of
the equation of state, initial conditions and the transport coefficients is
needed to completely solve non-ideal fluid dynamics. Once these are obtained
consistently one would then be in a position to do a thorough study of the
effects of dissipation in nuclear collisions.

In this section, we discuss basic aspects of numerical solution schemes for 
relativistic fluid dynamics. For the sake of simplicity, we consider 
the case of one conserved charge only. With the definitions Eqs.~(\ref{eq:cn}) -
(\ref{eq:P}) and the conservation laws Eqs.~(\ref{eq:Nmcons}) - (\ref{eq:Mcons}) 
the conservation equations can be solved numerically
with any scheme that also solves the non-relativistic conservation
equations. There is, however, one fundamental difference between the
non-relativistic equations and the relativistic ones.
In order to solve the latter
for $\cn,\, E,\, \tM$, the net charge density, energy density, and
momentum density in the calculational frame, one has to
know the equation of state $p(\eps,n)$ and $\tv$. The
equation of state, however, depends on $n,\, \eps$, 
the net charge density
and energy density in the local rest frame of the fluid. One therefore
has to locally transform from the calculational frame to the rest frame
of the fluid in order to extract $n,\, \eps, \, \tM$ from
$\cn,\, E,\, \tv$. In the non-relativistic limit, there is no difference
between $n$ and $\cn$, or $\eps$ and $E$ and the equation of state can
be employed directly in the conservation equations. Also, the momentum
density of the fluid is related to the fluid velocity by a simple
expression. The transformation between local rest frame and calculational frame
quantities is described explicitly below.

\subsection{Mathematical Structure of the equations of Non-ideal Fluid Dynamics}

In section \ref{sec:3+1} we formulated the equations of non-ideal fluid
dynamics for (1+1)-dimensional expansion in planar geometry, (1+1)-dimensional
cylindrical symmetric expansion ((1+1)-dimensional spherical symmetric 
expansion is discussed in Appendix \ref{append:sphsym}) , and (2+1)-dimensional
expansion with cylindrical symmetry. We now ask how to solve them. In this
connection it is instructive to count the number of the variables to be
determined and the number of equations available to determine them.

In non-ideal relativistic fluid dynamics we must find six fluid variables: $n$,
$\eps$, $p$ and three components of the ordinary velocity $\tv$; in addition we
must now find nine dissipative variables: the bulk pressure $\mPi$, the three
components of heat flow $\tq$, and the five independent (non-redundant)
components of the shear tensor $\tpi$. 
These fourteen variables are related by fourteen partial differential equations
governing the flow of relativistic non-ideal fluid: the five conservation
equations which are the net charge conservation equation, the energy
conservation equation and the three components of momentum conservation
equation; in addition we now have nine transport/relaxation equations for the
dissipative variables, namely the bulk pressure equation, the three components
of the heat flux equation and the five independent components of the shear
tensor equation.  In addition we have constitutive relations: we choose the
equation of state to be $p\equiv p(n,\eps)$.  

For (1+1)-dimensional expansions we need to determine only seven variables:  $n$,
$\eps$, $p$, $v$, $\mPi$, $q$, $\pi$ from a total of six differential equations
and an equation of state. The number of variables increases with the number of
spatial dimensions.  The system of fourteen unknowns and fourteen equations
(the 14-Field Theory) can be solved for the spatial variations of all unknowns
as a function of time once we are given a set of initial conditions that
specify the state and motion of the fluid at a particular time, plus a set of
boundary conditions where constraints are placed on the flow.

The techniques for solving these nonlinear equations are under investigation
and will be presented somewhere. Analytical methods can yield solutions for
some simplified problems, for example, (1+1)-dimensional flow with scaling
solution \cite{AM2}. But in general, this approach is too restrictive, and we
should recourse to numerical methods for a more realistic description. One of
the effective numerical methods of solving equations of fluid dynamics is to
replace the original differential equations by a set of finite difference
equations that determine the physical properties of the fluid on discrete space
and time meshes.

Given suitable initial and boundary conditions we follow the evolution of the
fluid by solving this discrete algebraic system at successive time steps. We
must make sure that the finite difference equations are numerically stable and
an efficient scheme must be found for handling shocks, which can produce
discontinuities in the solution or between mesh points.  We need numerical
techniques, which not only can handle the full nonlinear equations, but are
also versatile and flexible enough to (1) permit a detailed description of the
microphysics of the gas; (2) allow for structural complexities in the transient
regimes; (3) be generalized easily to include various processes of dissipation;
(4) account for departures from local thermodynamic equilibrium. This is a
subject of current investigation and will be presented somewhere.

\subsection{Transformation between Calculation Frame and Fluid Rest Frame}

In principle, the transformation is explicitly given by Eqs.~
(\ref{eq:cn}) -- (\ref{eq:P}), i.e., by finding the roots of a set of 14
nonlinear equations (the non-linearity enters through the equation of
state $p(\eps,n)$). In numerical applications, however, this
transformation has to be done several times in each time step and each cell.
It is therefore advisable to reduce the complexity of the transformation
problem. 
Boosting to a frame moving with velocity $-\tv$ (the collective velocity of the
fluid is $\tv$) the calculational and local rest frame transformations for the
heat flux and shear tensor are given by
\bea 
\pi^{\mu \nu } &=& \Lambda ^\m_\a \Lambda^\n_\beta \widehat{\pi}^{\a \b } ~,
\label{eq:pimntransf}\\
q^\m &=& \Lambda^\m_\a \widehat{q}^\a~,\label{eq:qtransf} 
\eea
where 
\bea
\Lambda^\m_\n &=& 
    \begin{pmatrix}  
    \g    &\quad \g \tv^{\rm T} \\
    \g \tv &\quad \tI +(\g -1)v^{-2} \tv \tv^{\rm T} 
    \end{pmatrix}  \label{eq:Lmunu}\;,\\
\widehat{q}^\m &=& (0,\cq^x,\cq^y,\cq^z) ~,\\
\widehat{\pi}^{\m\n} &=& 
    \begin{pmatrix}  
    0 &\quad  0     &\quad 0      &\quad 0 \\
    0 &\quad \t^{xx} &\quad \t^{xy} &\quad \t^{xz}  \\
    0 &\quad \t^{yx} &\quad \t^{yy} &\quad \t^{yz}\\ 
    0 &\quad \t^{zx} &\quad \t^{zy} &\quad \t^{zz}
    \end{pmatrix}  \label{eq:pimunuLRF}\;.
\eea
while the bulk pressure $\mPi$ transforms like the isotropic pressure $p$. In
Eq.~(\ref{eq:Lmunu}) $\tv$ is a $3\time 1$ column matrix, $\tv^{\rm T}$ is the
transpose of $\tv$ and $\tI$ is a $3\times 3$ identity matrix.

First note that unlike in relativistic ideal fluid dynamics $\tM$ and $\tv$ are
in general no longer parallel. In the ideal case $\tM$ and $\tv$ are
in parallel and  
\begin{equation}
\tM \cdot \tv \equiv M\, v = E- \eps \enspace ,
\end{equation}
where $M \equiv |\tM|$, $v \equiv |\tv|$.
One then obtains the relationship between $\eps$ and $E$ 
\begin{equation} \label{eq:epsilon}
\eps = E- M\, v ~,
\end{equation}
while the relationship between $n$ and $\cn$ 
\begin{equation} \label{eq:netcharge}
 n = \cn \sqrt{1-v^2}~,
\end{equation}
is the same as for non-ideal fluid dynamics because we are using the Eckart
definition of the 4-velocity.
With these equations $\eps$ and $n$ can be expressed in terms of $\cn,\,
E,\, M$ and $v$. The 5-dimensional root search is therefore reduced
to finding the modulus of $v$ for given $\cn,\, E,$ and $M$, which is a 
simple one-dimensional problem. To solve this, one uses the definition of
$M$, 
\begin{equation}
M=  (E+p) v \enspace .
\end{equation} 
This equation can be rewritten as a fixed point equation for $v$
for given $\cn,\, E,\, M$:
\begin{equation}\label{eq:veq}
v = \frac{M}{E+p \left(E-M\, v, \cn \sqrt{1-v^2}\right)} ~.
\end{equation}
The fixed point yields the modulus of the fluid velocity, from which
one can reconstruct $\tv = v \, \tM/M$, and find $\eps$ and
$n$ via Eq.~(\ref{eq:epsilon}). The equation of state $p(\eps, n)$ then
yields the final unknown variable, the pressure $p$.

Alternatively, one can use Eq.~(\ref{eq:veq}) to rewrite $\eps$ and $n$ as
\bea
\eps &=& E - {\tM^2\over E+p} ~,\\
n &=& \cn \left[1 - {\tM^2\over (E+p)^2}\right]^{1/2} = \cn\left[{\eps+p \over
E+p}\right]^{1/2}~,
\eea
from which we can get $\eps$ and $n$ by simultaneously solving
\bea
f(\eps,n) = (E-\eps)(E+p) - \tM^2 = 0 ~,\\
g(\eps,n) = (\cn^2-n^2)(E+p)^2 - \cn^2 \tM^2 = 0~,
\eea
and an equation of state $p(\eps, n)$ then
yields the final unknown variable, the pressure $p$.

However, in the non-ideal case dissipative effects brings complications to the
mathematical structure of the problem. But in some cases like the
(1+1)-dimensional expansions where the heat flux and the shear tensor can still
be replaced by some scalar quantities, as done in section \ref{sec:physprobs},
one can still generalize the above results. For example in (1+1) radial
expansion 
\bea
v_r &=& {M^r-\cq^r \over (E+\cp_r)} ~,\\
\eps &=& E - {{M^r}^2\over E+\cp_r} ~,\\
n &=& \cn \left[1 - {{M^r}^2\over (E+\cp_r)^2}\right]^{1/2}~,
\eea
from which we can get $\eps$ and $n$ by simultaneously solving
\bea
f(\eps,n) = (E-\eps)(E+\cp_r) - [{M^r}^2-{\cq^r}^2] = 0 ~,\\
g(\eps,n) = (\cn^2-n^2)(E+\cp_r)^2 - \cn^2 [M^r-\cq^r]^2 = 0 ~,
\eea
where $\cp_r = p(\eps,n)+\mPi+\t^{rr}$ and equation of state $p(\eps, n)$ then
yields the final unknown variable, the pressure $p$. In addition one needs the
transformations for the heat flux and the shear stress tensor which in this case
are given by 
\bea
q^r &=& \g \cq^r ~,\\
\pi^{rr} &=& \g^2 \t^{rr}~,
\eea
where the local rest frame heat flow is $\cq^r$ and the local rest frame shear
tensor is $\t^{rr}$.

The complications begin to arise when one moves away from the simple cases like
the one space dimensional expansions. In two and three space dimensional
expansions there are now more than one non-vanishing spatial components of
4-velocity flow. The dissipative fluxes, in particular the shear tensor,
strongly couples the various components of the non-vanishing spatial components
of the 4-velocity. Thus making the momentum components not to be in parallel
with the corresponding non-vanishing spatial components of the 4-velocity. 
In  finding the solution to this problem numerically the alternative method
discussed above for finding the local velocity, energy density and the net
charge density is no longer plausible. One then uses the generalized
first method. In this method the local velocities are calculated iteratively 
from Eqs.~(\ref{eq:tM}) and then the local energy density and net charge
density from Eqs.~(\ref{eq:eps}) and (\ref{eq:n}). In addition the
transformation of the heat flux and the shear tensor should be calculated from
Eqs.~(\ref{eq:pimntransf}) and (\ref{eq:qtransf}). But the latter transformation is
not necessary if the energy-momentum stress tensor is written in terms of the
calculational frame dissipative fluxes since their equations are already in this 
frame. In more complex situations this will be preferred to avoid further
difficulties. The calculational frame is then the natural frame
for the dissipative fluxes. In the nonrelativistic limit the equations should
reduce to the lab frame equations of non-ideal Newtonian fluids as shown in this
work (see Appendix~\ref{sec:non-rel}).

\subsection{Numerical Schemes}

To handle the coupled system of the partial differential equations presented 
here  we replace the differential equations by suitable discrete approximations
and solve these numerically. 
In general, to model a heavy-ion collision with non-ideal fluid dynamics
requires to solve the 5 conservation equations and the 9 transport/relaxation 
equations in three space dimensions. Since this is in general a formidable
numerical task, one should resorts  to the so-called operator splitting
method, i.e., the full 3-dimensional solution is constructed by solving
sequentially  three one-dimensional problems. 
More explicitly, all conservation equations are of the type
\begin{equation} \label{eq:dUcons}
\partial_t \, U + \sum_{i = x,y,z} \partial_i F_i(U) = 0 \enspace ,
\end{equation}
$U$ being $\cn,\, E,$ or $\tM$ and $F(U)$ being $\tN$ Eq.~(\ref{eq:cn}), $\tM$ 
Eq.~(\ref{eq:tM}) or $\tP$  Eq.~(\ref{eq:tP}). 
Such an equation is numerically solved on a space-time grid, 
and time and space derivatives are replaced by finite differences. 
The solution to the partial differential equation
(\ref{eq:dUcons}) in three space dimensions is obtained by solving
a sequence of one space dimension partial differential equations 
\begin{equation} \label{eq:dUconsi}
\partial_t \, U +  \partial_i F_i(U) = 0 \enspace ,
\end{equation}
where $i=x$. 
The 3-divergence operator in Eq.~(\ref{eq:dUcons}) is split into
a sequence of three partial derivative operators.
Physically speaking, in a given time step
one first propagates the fields in $x$ direction, then in $y$ direction, and
then in $z$ direction. In actual numerical applications it is
advisable to permutate the order $xyz$ to minimize systematical errors. 
The advantage of the operator splitting method is that there exists a
variety of numerical algorithms which solve evolution equations 
of the type Eq.~(\ref{eq:dUconsi}) in one space dimension (see, for instance,
\cite{RHLLE} and references therein). 

The transport/relaxation equations on the other hand are of convective type
\begin{equation} \label{eq:dUconv}
\partial_t \, U + \sum_{i = x,y,z} v_i\partial_i U = F_i(U) \enspace ,
\end{equation}
$U$ being $\mPi,\, \tq,$ or $\tpi$ and $F(U)$ being 
$(\mPi_E-\mPi)/\g\t_\mPi$, $(\tq_E-\tq)/\g\t_q$ or $(\tpi_E-\tpi)/\g\t_\pi$ in
the weak coupling limit of the causal transport/relaxation
equations. 
These equations are of non-conservative convective type. They should be 
solved by numerical schemes which solves the advective equations of the form
\be
\PD_t U + \tv.\tnabla U = \mbox{Sources}
\ee
rather than the conservative form
\be
\PD_t U + \tnabla \{U \tv\} = \mbox{Sources} \label{eq:conservative}
\ee
unless the transport equations are recast in the form of
Eq.~(\ref{eq:conservative}). 
The actual numerical schemes for future calculations are a subject of current
investigation and will be presented somewhere.

\section{Summary and Discussion}
\label{sec:summary}
%
We have provided a complete set of equations for relativistic dissipative  
hydrodynamics in their 3+1 representation. Causality has been accounted for by
using the extended causal description  of thermodynamics, relativistically
formulated on a phenomenological level. In
contrast to the conventionally used compressible Navier-Stokes description of
non-ideal hydrodynamics, the equations of extended causal thermodynamics
guarantee finite propagation speeds of heat and viscous signals and yield
stable local thermodynamic equilibria. 

The extended theories provide a model for evolving dissipative fluids in a
causal and stable manner. In addition, these theories allow one to describe
physical systems which cannot be modeled by Navier-Stokes-Fourier, in
particular situations where the relaxation times are not equal to the
microscopic collision time.

In their simplest form (Maxwell-Cattaneo limit) these transport equations
describe the relaxation on finite timescales of dissipative fluxes towards the
standard Navier-Stokes-Fourier equations, which guarantees finite signal
propagation.

A causality preserving formulation is required whenever the thermodynamic
timescale becomes comparable to the dynamical timescale and therefore the
assumption of local thermodynamic equilibrium is not justified. In many
problems of interest the inertia due to the dissipative contributions to the
energy-momentum stress tensor can be neglected. The 3+1 formulation of the
corresponding simplified set of equations is given.

The five conservation laws for particle number, energy and momentum, and the
nine evolution equations for the thermodynamic fluxes (i.e., the 14-Fields
Theory of Relativistic Non-ideal Fluid Dynamics) form a hyperbolic system of
first order partial differential equations tractable by numerical methods. 

Ideally one would like to be able to extract the transport coefficients and
associated time and length scales from a particular model of an equation of
state (interactions among the constituents). Then study the effects of
dissipative nonequilibrium processes on the space-time evolution of the system
and on the calculated distributions of particles. And finally one would like to
compare the predicted distributions with those observed in experiments.

Finally it is noted that non-ideal fluid dynamics may provide a better way of
combining hydrodynamic approach and the kinetic (sequential scatering) approach
required for a realistic description of high enrgy nuclear collisions.  

\appendix
%
\section{ Spherically symmetric flow: fireball expansion}
\label{append:sphsym}
For a one-dimensional, spherical symmetric flow the terms in $(\PD/\PD \q)$ and
$(\PD/\PD \f)$ vanish identically. 
When viewed by an observer moving relative to the new coordinates with proper
velocity $v_r$ in the radial direction, the physical content of space consists
of an anisotropic fluid of energy density $\eps$, radial pressure $\cp_r$,
tangential pressure $\cp_\perp$, radial heat flux $q$. Thus when viewed by this
comoving observer the covariant net charge 4-current and energy-momentum tensor in Minkowski coordinates is
\bea
\hat{N}^\m  &=& (n,0,0,0) ~,\\ 
\hat{T}^{\m\n} &=& \left( \begin{array}{cccc} 
         \eps   & q     &    0      & 0	 \\
         q     & \cp_r  &    0      & 0  \\  
         0     & 0     & \cp_\perp  & 0   \\
         0     & 0     & 0         & \cp_\perp
    \end{array} \right) ~. \label{eq:TmunuLRF3}
\eea
Then a Lorentz transformation leads to
\bea
N^\m &=& n u^\m~,\\
T^\m_\n &=& \eps u^\m u_\n -\cp_{\rm eff}\Delta^\m_\n +
\left(\cp_r-\cp_\perp\right)\left[m^\m m_\n + \3 \Delta^\m_\n\right] +
q\left(m^\m u_\n + m_\n u^\m\right)~, \label{eq:Tmunu-calc3}
\eea
with $\Delta^\m_\n = \delta^\m_\n - u^\m u_\n$, $\cp_{\rm eff} =
\3(\cp_r+2 \cp_\perp)$, $\cp_r = p+\mPi+\pi$, $\cp_\perp=p+\mPi-\pi/2$, 
$u^\m = (\g, \g v_r,0,0)$ and  $m^\m = (\g v_r, \g, 0,0)$. The heat 4-current 
 and the pressure tensor can be written respectively as 
\bea
q^\m &=& q m^\m ,~ \\
P^\m_\n &=& -\cp_{\rm eff}\Delta^\m_\n +
\left(\cp_r-\cp_\perp\right)\left[m^\m m_\n + \3 \Delta^\m_\n\right]~.
\eea
The local rest frame net charge 4-current and the energy-momentum stress tensor
are obtained by boosting back with  $u^\m (t,r,\f,\q) = \widehat{u}^\m
(t,r,\f,\q) =
(1,0,0,0)$ and $m^\m (t,r,\f,\q) = \widehat{m}^\m (t,r,\f,\q) = (0,1,0,0)$.
The nonvanishing components of the net charge and the energy-momentum tensor
\bea
N^0 &=& n\g \sks N^r = n\g v_r~,\\
T^0_0 &=& \cw \g^2 -\cp_r+ 2 q \g^2 v_r = T^{00} ~,\\
-T^0_r &=& \cw \g^2 v_r + q \g^2 (1+ v_r^2) = T^r_0=T^{0r}~,\\
-T^r_r &=& \cw \g^2 v_r^2 +\cp_r + 2 q \g^2 v_r = T^{rr} ~,\\
-T^\f_\f &=& -T^\q_\q = \cp_\perp~,\\ 
\eea
with $\cw \equiv \eps+\cp_r$. 
The local velocity, energy density and net charge
density are obtained from
\be
N^0 = \g n \sks \eps = T^{00} -(T^{0r}+q)v_r \sks T^{0r} = (T^{00}+\cp_r) v_r +
q~,\\ 
\ee
that is 
\bea
v_r &=& {T^{0r}-q \over T^{00}+\cp_r} ~,\\
\eps &=& T^{00} - {{T^{0r}}^2- q^2 \over T^{00}+\cp_r} ~,\\
n &=& (1-v_r^2)^{1/2}N^0 ~.
\eea
The net charge conservation $\PD_\m N^\m = 0$ and the energy-momentum
conservation $\PD_\m T^\m_\n =0$ can be written as
\bea
\PD_\m N^\m \equiv 0 &\Lra& \PD_t N^0 +{1\over r^\a}\PD_r\{r^\a N^0 v_r\} =0
~,\\
\PD_\m T^\m_0 \equiv 0 &\Lra& \PD_t T^{00} +{1\over r^\a}\PD_r\{r^\a T^{r0}\}
=0 ~,\\
\PD_\m T^\m_r \equiv 0 &\Lra& \PD_t T^{0r} +{1\over r^\a}\PD_r\{r^\a T^{rr}\}
-{\a\over r} T^\f_\f=0~,
\eea
where $\a=2$ for spherical geometry. For numerical purposes we recast the above
equation as follows
\bea
&&\PD_t N^0 + \PD_r\{N^0 v_r\} = {\a \over r}  N^0 v_r ~,\\
&&\PD_t T^{00} + \PD_r\{(T^{00}+\cp_r) v_r +q\} =-\a{v_r\over r} (T^{00} +\cp_r)
-\a {q\over r}~,\\
&&\PD_t T^{0r} + \PD_r\{(T^{0r}+q) v_r+\cp_r\} = -\a {v_r\over r} (T^{0r} +q)
-{\a\over r}(\cp_r - \cp_\perp)~,
\eea
The structure of these equations is similar to the one dimensional case with
additional geometrical terms.

For the transport equations recall that there is only one independent
component  of the heat flux and of the shear stress tensor. For the heat flux
this translates into the equation for $q$, the heat flow. For the shear stress
tensor we choose either 
the $\f\f$ or the $\q\q$ components. This is just for
convenience since these choices makes the calculations more tractable. The
simplicity comes from the property that these transverse components of the
shear stress tensor do not change under the boost and hence they do not pick up
the velocity or the Lorentz gamma factors which would otherwise render the
solution more difficult. Thus we see that in (1+1)-dimensional spherical 
symmetric flows all the dissipative fluxes can be represented by scalar
quantities.  
In the case in which there is no viscous/heat couplings ($\a_i$'s and $l_A$'s
are zero)  the weak coupling limit of the causal transport equations in the
cylindrical and fireball geometry takes the form
\bea
\t_\mPi\dot{\mPi} +\mPi &=& -\z\mTheta -{1\over 2}\mPi\left(\t_\mPi\mTheta+\z
T\left({\t_\mPi\over \z T}\right)^{\dot{}} \right) ~,\\
\t_q\dot{q} + q &=& -\k T\left({T^\pr\over T} + a\right) 
-{1\over 2} q\left(\t_q\mTheta - \kappa T^2\left({\t_q\over \kappa
T^2}\right)^{\dot{}}\right) ~,\\
\t_\pi\dot{\pi} +\pi &=& -2\cdot 2\h \s  
-{1\over 2}\pi\left(\t_\pi\mTheta+\h T\left(\t_\pi\over\h
T\right)^{\dot{}}\right)~,
\eea
where
\bea
\s &=& \left[ -{\g v_r\over r} +{1\over 3}\mTheta \right]~,\\
\dot{f} &\equiv& \left[\g{\PD \over \PD t}+\g v_r {\PD \over \PD
r}\right] f~,\\
f^{\pr} &\equiv& \g\left[\g v_r{\PD \over \PD t}+\g {\PD \over \PD
r}\right] f~,\\
\mTheta &\equiv& \theta + \alpha {\g v_r\over r}~,\\
\theta &\equiv&  {\PD \over \PD t}\gamma +{\ppr} \gamma v_r \,,
\eea
The relationship between the 
calculational frame and the local rest frame components of
the shear tensor can be read off from Eqs.~(\ref{eq:Tmunu-calc3}) and
(\ref{eq:TmunuLRF3}). In particular
\bea
q^r &=& \g \cq^r =\g q~,\\
\pi^{rr} &=& \g^2 \t^{rr}=\g^2 \pi~,\\
\pi^\q_\q &=& \t^{\q\q} = -{\pi\over 2}~,\\
\pi^\f_\f &=& \t^{\f\f} = -{\pi\over 2}~ .
\eea
Including the bulk equation we only have three transport equations.  Again
these equations have the same structure as in one--dimensional case, except for
geometrical terms.

From the energy-momentum conservation, $\PD_\m T^\m_\n =0$, in the calculational
frame we have
\bea
&&u^\m\PD_\m \eps + (\eps+ \cp_{\rm eff})\mTheta +\PD_\m q^\m =
(\cp_r-\cp_\perp)\left[m_\m m_\n+\3\Delta_{\m\n}\right]\s^{\m\n} + qa^\m
m_\m~,\label{eq:e-lrf}\\
&&(\eps+\cp_{\rm eff}) a^\m +\Delta^{\m\n}\left(m_\n u^\a\PD_\a q + q u^\a\PD_\a
m_\n - \PD_\n \cp_{\rm eff} +\PD_\b \left\{(\cp_r-\cp_\perp)\left[m^\b m_\n + \3
\Delta^\b_\n\right]\right\}\right) \nonumber\\
&&+ q m_\n \s^{\m\n} +{4\over 3}\mTheta q m^\m = 0 ~,\label{eq:m-lrf}
\eea
which are the local energy density conservation equation, Eq.~(\ref{eq:e-lrf}),  
and the local rest frame equations of motion, Eqs.~(\ref{eq:m-lrf}). 
Contracting the second equation with $m_\m$ we get
\be
m^\m\PD_\m \cp_r + (\cp_r-\cp_\perp)\PD_\m m^\m - (\eps+\cp_\perp) a_\m m^\m +
{4\over 3} \mTheta q + u^\m \PD_\m q - q m^\m m^\n \s_{\m\n} = 0 ~,
\ee
where
\bea
\s^{\m\n} &=& {1\over 2} \s\left(m^\m m^\n +\3\Delta^{\m\n}\right) ~,\\
\s &=& -\left[2\theta -2{\g v_r\over r}\right] 
\Longrightarrow {1\over 2}\s + \mTheta = 3 {\g v_r\over r}~.
\eea
Explicitly, net charge density, the energy and the radial pressure equations
are
\bea
&&\dot{n} + n\mTheta = 0~,\\
&&\dot{ \eps} + (\eps+ \cp_{\rm eff})\mTheta +{1\over \g} q^{\prime} =
-\3 (\cp_r-\cp_\perp)\s - 2 q {a\over \g}~,\label{eq:e-loc}\\
&&\cp_r^{\prime} +\g \dot{q} = -(\eps+\cp_\perp) a - (\cp_r-\cp_\perp)\g 
\ca -{2\over 3}\left[2\mTheta +\s\right]\g q ~,
\eea
where
\bea
a &=& {\g\over v_r}\left[\PD_t\g + v_r^2\PD_r \g v_r \right] ~,\\
\ca &=& {1\over \g}\left(a+ 2{\g\over r}\right)~,\\
\dot{f} &\equiv& \g\left[\PD_t + v_r \PD_r\right] f~,\\
f^{\prime} &\equiv& \g^2 \left[v_r\PD_t + \PD_r\right] f~,
\eea
\section{Physical Components and Transformation Matrices}
\label{append:physcomp}
Although the covariant and contravariant components of the vectors (e.g, 4-velocity) and tensors
(energy-momentum tensor) in curved space coordinates (e.g., cylindrical and
spherical coordinates) are useful for calculations the normalizations brought
in by the metric make them inconvenient for physical interpretations. More
convenient are components on orthonormal tetrads carried by the fluid elements.
The line element is given by 
\be
ds^2 = g_{\m\n} d x^\m d x^\n  ~.
\ee
Since the metric is diagonal we will denote the nonvanishing components by 
$g^{(\n)(\n)}$ and respectively $g_{(\n)(\n)}$ for its inverse. 
For a vector $A^\n$, with the abstract
contravariant component $A^{(\n)}$ we must associate the {\em physical component} 
\be
A(\n) = |g_{(\n)(\n)}|^{1/2} A^{\n}~,\label{eq:vcon}
\ee
with no summation implied.
For covariant components the physical
components are related to covariant components by the expression
\be
A{(\n)} = |g^{(\n)(\n)}|^{1/2} A_{\n}~,\label{eq:vco}
\ee
with no summation implied. 
To compute the physical components of a tensor the
{\em physical components} of $T^{\m\n}$ in terms of its covariant components are
\be
T(\m,\n) = |g^{(\m)(\m)}|^{1/2}|g^{(\n)(\n)}|^{1/2}T_{\m\n} ~.\label{eq:tco}
\ee
Similarly for contravariant components we have
\be
T(\m,\n) =|g_{(\m)(\m)}|^{1/2}|g_{(\n)(\n)}|^{1/2}T^{\m\n} ~.\label{eq:tcon}
\ee
The parenthetic index has no tensorial significance being merely a label. 
The physical components do not of course transform as tensors, but their
transformation law can be easily deduced.
One notices that the diagonal elements of a mixed second rank tensor are the
same as the physical components.
For mixed tensors the physical components are related to the abstract components
by
\be
T(\m,\n) = \left|{g_{(\m)(\m)}\over g_{(\n)(\n)}}\right|^{1/2} T^\m_\n 
= \left|{g_{(\n)(\n)}\over g_{(\m)(\m)}}\right|^{1/2} T^\n_\m
\ee
or
\be
T(\m,\n) = \left|{g_{(\m)(\m)}\over g_{(\n)(\n)}}\right|^{1/2} g^{\m\a}T_{\a\n}
= \left|{g_{(\m)(\m)}\over g_{(\n)(\n)}}\right|^{1/2} g_{\n\a} T^{\a\m}
\ee

\begin{itemize}

\item {\em Spherical symmetry}

The metric and the line element in spherical coordinates read
\bea
g^{\m\n} &=& \mbox{diag} (1,-1,-r^{-2},-r^{-2}\sin^{-2}\f) \sks
g_{\m\n} = \mbox{diag} (1,-1,-r^2,-r^2\sin^2\f)~,\\
d s^2 &=& dt^2 - dr^2 -r^2 d\f^2 - r^2 \sin^2\f d\q^2.
\eea
and the determinant $g= -r^4\sin\f$ while the transformation matrix for the
derivatives is
\be
\left( \begin{array}{c}
\displaystyle{\frac{\PD}{\PD t}} \\\\ \displaystyle{\frac{\PD}{\PD x}}
\\\\
\displaystyle{\frac{\PD}{\PD y}} \\\\ \displaystyle{\frac{\PD}{\PD z}} \end{array} \right) =
\left( \begin{array}{cccc} 
         1 & 0               &0                 &0	 \\\\
         0 & \sin\f \cos\q   & \cos\f \cos\q    &-\sin\q  \\\\  
         0 & \sin\f\sin \q & \cos\f \sin \q   & \cos\q   \\\\
         0 & \cos\f          & -\sin\f          & 0
    \end{array} \right)  
    \left( \begin{array}{c}
       \displaystyle{\frac{\PD}{\PD t}} \\\\ 
       \displaystyle{\frac{\PD}{\PD r}} \\\\
       \displaystyle{\frac{1}{r}\frac{\PD}{\PD \f}} \\\\ 
       \displaystyle{\frac{1}{r \sin\f} \frac{\PD}{\PD \q}}
       \end{array} \right) \label{eq:sphtransf}~.
\ee

The relations between the abstract and physical
components of the 4-velocity in spherical
coordinates read (respectively contravariant and covariant)
\bea
u^{(0)} &=& u_t\sks u^{(1)} = u_r \sks u^{(2)} = u_\f/r \sks u^{(3)} =
u_\q/(r\sin\f)~.\\
u_0 &=& u_t\sks u_1 = u_r \sks u_2 = r u_\f \sks u_3 = r\sin\f\, u_\q~.
\eea
For a symmetric tensor in spherical coordinates we have, for contravariant
components, the relations between physical and abstract components
\bea
T^{00}&=& T_{tt} \sks T^{11} = T_{rr} \sks T^{12} = T_{r\f}/r 
\sks T^{13} = T_{r\q}/(r\sin\f) ~,\\
T^{22} &=& T_{\f\f}/r^2 \sks T^{23} = T_{\f\q}/( r^2 \sin\f) \sks T^{33} =
T_{\q\q}/(r\sin\f)^2~,
\eea
with analogous expressions for covariant components.

\item {\em Cylindrical symmetry}

The the metric and the line element in cylindrical coordinates read
\bea
g^{\m\n} &=& \mbox{diag} (1,-1,-r^{-2},-1) \sks
g_{\m\n} = \mbox{diag} (1,-1,-r^2,-1)~,\\
d s^2 &=& dt^2 - dr^2 -r^2 d\f^2 - d z^2~.
\eea
and the determinant $g= r^2$ while the transformation of the
derivatives is
\be
\left( \begin{array}{c}
\displaystyle{\frac{\PD}{\PD t}} \\\\ \displaystyle{\frac{\PD}{\PD x}}
\\\\
\displaystyle{\frac{\PD}{\PD y}} \\\\ \displaystyle{\frac{\PD}{\PD z}} \end{array} \right) =
\left( \begin{array}{cccc} 
         1 & 0               &0                 &0	 \\\\
         0 & \cos\f          & -\sin\f          &0       \\\\  
         0 & \sin \f         & \cos\f           &0        \\\\
         0 & 0               &0                 &1
    \end{array} \right)  
    \left( \begin{array}{c}
       \displaystyle{\frac{\PD}{\PD t}} \\\\ 
       \displaystyle{\frac{\PD}{\PD r}} \\\\
       \displaystyle{\frac{1}{r}\frac{\PD}{\PD \f}} \\\\ 
       \displaystyle{\frac{\PD}{\PD z}}
       \end{array} \right) \label{eq:cyltransf}~.
\ee
The relations between the abstract and physical
components of the 4-velocity in cylindrical coordinates are
(respectively for contravariant and covariant components)
\bea
u^{(0)}&=& u_t \sks u^{(1)} = u_r \sks u^{(2)} = u_\f/r \sks u^{(3)} = u_z~.\\
u_0&=& u_t \sks u_1 = u_r \sks u_2 = r u_\f \sks u_3 = u_z~.
\eea
For a symmetric tensor in cylindrical coordinates we have (for contravariant
components) 
\bea
T^{00} &=& T_{tt} \sks T^{11} = T_{rr} \sks T^{12} = T_{r\f}/r \sks
T^{13} = T_{rz} ~,\\
T^{22} &=& T_{\f\f}/r^2  \sks T^{23} = T_{\f z}/r \sks  T^{33} = T_{z z}~,
\eea
with analogous expressions for covariant components.
\end{itemize}

\section{Non-ideal fluid dynamics in the non-relativistic limit}
\label{sec:non-rel}
In Cartesian coordinates $(t,x,y,z)$ the governing equations can be written
compactly as a single equation as 
\be
{\partial U \over \partial t}+ {\partial F\over \partial x} + \
{\partial G \over \partial y}+ {\partial H \over \partial z}= 0
\ee
where
\bea
U &\equiv& \left[ \begin{array}{c}
\rho\\ \cm^x \\ \cm^y \\ \cm^z \\ E
\end{array} \right] 
\sks 
F \equiv \left[ \begin{array}{c}
\rho v_x\\ \cm^x v_x + \cp_x\\ \cm^y v_x  + \pi^{xy}\\
\cm^z v_x +\pi^{xz}\\ (E+\cp_x) v_x + \pi^{yx} v_y + \pi^{zx} v_z
+\cq^x \end{array}\right]  \nonumber\\
G &\equiv& \left[ \begin{array}{c}
\rho v_y\\ \cm^x v_y+\pi^{yx}\\ \cm^y v_y + \cp_y\\ 
\cm^z v_y + \pi^{yz}\\
 (E+\cp_y) v_y + \pi^{xy} v_x + \pi^{zy} v_z
+\cq^y \end{array}\right]  \nonumber\\
H &\equiv& \left[ \begin{array}{c}
\rho v_z\\ \cm^x v_z+\pi^{zx}\\ \cm^y v_z+\pi^{zy}\\
\cm^z v_z + \cp_z\\ 
 (E+\cp_z) v_z + \pi^{xz} v_x + \pi^{yz} v_y
+\cq^z \end{array}\right]  
\eea
in which $\cm^i \equiv \rho v_i$ and $\cp_i \equiv p+\mPi+\pi^{ij}\delta_{ij}$. 
The dissipative fluxes are governed by the following relaxational transport equation 
\be
{\partial U \over \partial t}+ v_x{\partial U\over \partial x} + \
v_y{\partial U \over \partial y}+ v_z{\partial U \over \partial z}= -{1\over
\t_U}\left[U+ F\right]
\ee
where $\t_U = \{\t_\Pi,~\t_q,~\t_\pi\}$ and 
\be
U \equiv \left[ \begin{array}{c}
\mPi\\ \cq^x\\ \cq^y \\ \cq^z \\ \p^{xx} \\ \pi^{yy} \\ \pi^{zz} \\
\pi^{xy}=\pi^{yx}\\ \pi^{yz}=\pi^{zy}\\ \pi^{zx}=\pi^{xz}
\end{array} \right]  \sks
F \equiv \left[ \begin{array}{c}
-\z\mTheta\\ -\k{\partial T\over \partial x}\\ -\k{\partial T\over \partial y} \\ 
-\k{\partial T\over \partial z} \\ 
-2\h\left({\partial v_x\over \partial x} - {1\over 3}\mTheta\right)\\ 
-2\h\left({\partial v_y\over \partial y} - {1\over 3}\mTheta\right)\\ 
-2\h\left({\partial v_z\over \partial z} - {1\over 3}\mTheta\right) \\
-\h\left[{\partial v_y\over \partial x}+{\partial v_x\over \partial y}\right]\\ 
-\h\left[{\partial v_z\over \partial y}+{\partial v_y\over \partial z}\right]\\ 
-\h\left[{\partial v_x\over \partial z}+{\partial v_z\over \partial x}\right]
\end{array} \right] 
\ee
in which
\be
\mTheta = {\partial v_x \over \partial x} + {\partial v_y \over \partial y}
+{\partial v_z \over \partial z}
\ee

For one dimensional Cartesian coordinates $(t,z)$, pure cylindrical coordinates 
$(t,r,\phi, z)$ with $v_\phi = v_z = 0$ and 
pure spherical coordinates $(t,r,\phi,\theta)$ with $v_\phi =v_\theta =0$ the
governing equations can be written in one single equation
\be
{\partial U \over \partial t} 
+ {1\over r^{\a-1}}{\partial \over \partial r}\left\{r^{\a-1} U v\right\}  
+ {1\over r^{\a-1}}{\partial \over \partial r}\left\{r^{\a-1}F \right\} 
+ {\partial G \over \partial r}
+{(\a-1)\over r }H = 0
\ee
where $\a=1,2,3$ for Cartesian, cylindrical and spherical coordinates
respectively and in particular for cylindrical coordinates
\bea
U &\equiv& \left[ \begin{array}{c}
\rho\\ \cm^r \\ E
\end{array} \right]  
\sks 
F \equiv \left[ \begin{array}{c}
0\\ \pi^{rr} \\ \cp v_r + \cq^r \end{array}\right] 
\sks
G \equiv \left[ \begin{array}{c}
0\\ \cp\\ 0 \end{array}\right] \sks
H \equiv \left[ \begin{array}{c}
0\\ -\pi^{\f\f} \\ 0 \end{array}\right] 
\eea
The governing equations for the dissipative fluxes can be written in a single
compact equation
\be
{\partial U \over \partial t} 
+ v_r{\partial U\over \partial r}  
= -{1\over \t_U}\left[U+F(U)\right] 
\ee
where in cylindrical coordinates
\be
U \equiv \left[ \begin{array}{c}
\mPi \\ \cq^r \\ \pi^{rr}\\ \pi^{\f\f}\\\pi^{zz}
\end{array} \right]  
\sks 
F \equiv \left[ \begin{array}{c}
-\z\mTheta\\
-\k {\partial T\over \partial r} \\
-2\h\left({\partial v_r\over \partial r}-{1\over 3}\mTheta\right)\\
-2\h\left({v_r\over r} - {1\over 3}\mTheta\right)\\
 {2\over 3}\h\mTheta
\end{array}\right] 
\ee
and in spherical coordinates
\be
U \equiv \left[ \begin{array}{c}
\mPi \\ \cq^r \\ \pi^{rr}\\ \pi^{\f\f}\\\pi^{\q\q}
\end{array} \right]  
\sks 
F \equiv \left[ \begin{array}{c}
-\z\mTheta\\
-\k {\partial T\over \partial r} \\
-2\h\left({\partial v_r\over \partial r}-{1\over 3}\mTheta\right)\\
-2\h\left({v_r\over r} - {1\over 3}\mTheta\right)\\
-2\h\left({v_r\over r} - {1\over 3}\mTheta\right) 
\end{array}\right] 
\ee
while in Cartesian coordinates
\be
U \equiv \left[ \begin{array}{c}
\mPi \\ \cq^z \\ \pi^{zz}\\ \pi^{yy}\\\pi^{xx}
\end{array} \right]  
\sks 
F \equiv \left[ \begin{array}{c}
-\z\mTheta\\
-\k {\partial T\over \partial z} \\
-2\h\left({\partial v_z\over \partial z}-{1\over 3}\mTheta\right)\\
{2\over 3}\h\mTheta\\
{2\over 3}\h\mTheta 
\end{array}\right] 
\ee
in which 
\be
\mTheta = {1\over r^{\a-1}} {\partial \over \partial r} \left(r^{\a-1} v_r\right)
\ee
Note that in the Cartesian coordinates ($\a=1$) the subscript/superscript $r$
is replaced by $z$ and the differentiation with respect to $r$ becomes
differentiation with respect to $z$. Also the transverse components of
$\pi^{\m\n}$, namely $\pi^{xx}$ and $\pi^{yy}$ do not vanish and they are equal
to $-\pi^{zz}/2$.  In the spherical ($\a=3$) case  the transverse components
are the $\pi^{\f\f}$ and the $\pi^{\q\q}$ and each equals $-\pi^{rr}/2$. In the
cylindrical coordinates, $\a=2$, the transverse components of $\pi^{\m\n}$ are
$\pi^{\f\f}$ and $\pi^{zz}$.
\acknowledgments
%
I would like to thank Rory Adams, Tomoi Koide and Pasi Huovinen for reading 
the manuscript and for valuable comments.

%

%


\begin{thebibliography}{}
%
%
\bibitem{AMII} A. Muronga, nucl-th/0611091.
%
\bibitem{HS} H. Stoecker, E. L. Bratkovskaya, M. Bleischer, S. soff and X. Zhu,
 nucl-th/0412022. 
%
\bibitem{AM2} A. Muronga, Phys. Rev. {\bf C69} (2004) 044901.
%
\bibitem{MR} A. Muronga and D.H. Rischke, nucl-th/0407114.
%
\bibitem{DT} D. Teaney, Phys. Rev. {\bf C68}, 034913 (2003).
%
\bibitem{AG} M. A. Aziz and S. Gavin, Phys. Rev. {\bf C70}, 034905 (2004).
%
\bibitem{KP}  K. Paech, H. Stocker and  A. Dumitru, 
               Phys. Rev. {\bf C68} (2003) 044907.
%
\bibitem{PD} C. Pujol and D. Davesne, Phys. Rev. C67 (2003) 014901.
%
\bibitem{HiGy} T. Hirano and M. Gyulassy, nucl-th/050649.
%
\bibitem{HSC} U. Heinz, H. Song and A.K. Chaudhuri, Phys. Rev. {\bf C73}, 034904
(2006).
%
\bibitem{TK} T. Koide, G. S Denicol, Ph. Mota and T. Kodama, hep-ph/0609117.
%
\bibitem{BRW}  R. Baier, P. Romatschke and  U. A. Wiedemann, 
Phys. Rev. C73 (2006) 064903; nucl-th/0604006.
%
\bibitem{BR}  R. Baier and  P. Romatschke, nucl-th/0610108.
%
\bibitem{DHR} D.H. Rischke, S. Bernard and  Joachim A. Maruhn, 
              Nucl. Phys. {\bf A 595} (1995) 346. 
%
\bibitem{CE} C. Eckart, Phys. Rev. {\bf 58}, (1940) 919.
%
\bibitem{LL} L.D. Landau and E.M. Lifshitz {\it Fluid Mechanics}
(London: Perganon), 1959.
%
\bibitem{HL} W.A. Hiscock and L. Lindblom, Ann. Phys. {\bf 151}, (1983) 466.
%
\bibitem{HG} H. Grad, Commun. Pure Appl. Math. {\bf 2} (1949)
331.
%
\bibitem{IM} I. M\"uller, Z. Phys. {\bf 198}, (1967) 329.
%
\bibitem{IS} W. Israel and J.M. Stewart, Ann. Phys. {\bf 118}, (1979) 341.
%
\bibitem{LMR} I Liu, I. M\"uller and T. Ruggeri, Ann. Phys. {\bf 169}, (1986)
191.
%
\bibitem{Bjorken}
  J.D. Bjorken, \jou{\PRD}{27}{140}{1983}. 
%
\bibitem{RHLLE} V. Schneider et al., J. Comput. Phys. {\bf 105} (1993) 92.
%
	       
\end{thebibliography}
\end{document}